\documentclass[twocolumn]{aastex62}
\usepackage{natbib}
\usepackage{amsmath}

\newcommand{\angstrom}{\textup{\angstrom}}

\newcommand\effco{\epsilon_{\rm ff, co}}
\newcommand\effst{\epsilon_{\rm ff, st}}
\newcommand\effint{\epsilon_{\rm ff, int}}

\newcommand\tff{\tau_{\rm ff}}
\newcommand\tffi{\tau_{\rm ff, 0}}

\newcommand\rhost{\rho_{\rm star}}
\newcommand\rhoco{\rho_{\rm core}}
\newcommand\rhog{\rho_{\rm gas}}

\newcommand\tret{t_{return}} 
\newcommand\fret{f_{return}}

\newcommand\Ms{M_{\odot}}

\newcommand\MsMy{M_{\odot} \cdot Myr^{-1}}
\newcommand\Mspp{M_{\odot} \cdot pc^{-2}}
\newcommand\Msppp{M_{\odot} \cdot pc^{-3}}
\newcommand\cc{cm^{-3}}

\newcommand\stf{star formation }
\newcommand\sfing {star-forming }
\newcommand\cof{core formation }
\newcommand\gc{globular cluster }

\newcommand\gcs{globular clusters }

\newcommand\icm{intra-cluster medium }

\newcommand\sfe{star formation efficiency }
\newcommand\cofe{core formation efficiency }

\newcommand\sfr{star formation rate }
\newcommand\cofr{core formation rate }

\newcommand\fft{free-fall time }

\newcommand\msps{multiple stellar populations }
\newcommand\cm{color-magnitude }

\newcommand\hhb{hot hydrogen burning }
\newcommand\smss{supermassive stars }

\accepted{2021 October 23}

\shorttitle{Star Formation Bouncing Back with Rebounding Cores}
\shortauthors{Parmentier \& Pasquali}

\begin{document}

\title{Rebounding Cores to Build Star Cluster Multiple Populations}

\correspondingauthor{Genevieve Parmentier}
\email{gparm@ari.uni-heidelberg.de}

\author[0000-0002-2152-4298]{G. Parmentier}
\email{gparm@ari.uni-heidelberg.de}
\affiliation{Astronomisches Rechen-Institut, Zentrum f\"ur Astronomie der Universit\"at Heidelberg, M\"onchhofstr. 12-14, D-69120 Heidelberg, Germany}

\author{A. Pasquali}
\affiliation{Astronomisches Rechen-Institut, Zentrum f\"ur Astronomie der Universit\"at Heidelberg, M\"onchhofstr. 12-14, D-69120 Heidelberg, Germany}



\begin{abstract}
We present a novel approach to the riddle of star cluster multiple populations.  Stars form from molecular cores.  But not all cores form stars.  Following their initial compression, such 'failed' cores re-expand, rather than collapsing.  We propose that their formation and subsequent dispersal regulate the gas density of cluster-forming clumps and, therefore, their core and star formation rates.  Clumps for which failed cores are the dominant core type experience star formation histories with peaks and troughs.  In contrast, too few failed cores results in smoothly decreasing star formation rates.  We identify three main parameters shaping the star formation history of a clump: the star and core formation efficiencies per free-fall time, and the time-scale on which failed cores return to the clump gas.  The clump mass acts as a scaling factor.  We use our model to constrain the density and mass of the Orion Nebula Cluster progenitor clump, and to caution that the  star formation histories of starburst clusters may contain close-by peaks concealed by stellar age uncertainties. 
Our model generates a great variety of star formation histories.  Intriguingly, the chromosome maps and O-Na anti-correlations of old globular clusters  also present diverse morphologies.  This prompts us to discuss our model in the context of globular cluster multiple stellar populations.  More massive globular clusters exhibit stronger multiple stellar population patterns, which our model can explain if the formation of the polluting stars requires a given stellar mass threshold.\end{abstract}


\section{Introduction} \label{sec:intro}

The presence of \msps in star clusters remains a great riddle.  Increasingly complex patterns have been revealed in the color-magnitude diagrams of Galactic globular clusters, which were once thought to epitomise simple stellar populations \citep[see e.g.][]{mil08,and09,mil15}.  These patterns stem from star-to-star abundance variations in He \citep{pio07}, and in C, N and O \citep{mil15}.  CNO abundance variations in individual \gcs had already been identified by earlier spectroscopic studies \citep[e.g.][]{kra94}, and Na-Mg-Al abundance variations are also common \citep[][and references there in]{kra94,car06,cha16}.  These abundances are either correlated (Na-N-He) or anti-correlated (Na-O, N-C), which is the signature of hot-hydrogen burning \citep[i.e. hydrogen-burning through the CNO-cycle and the NeNa- and MgAl-chains, ][]{lan93}.  The current consensus is that the low-mass stars enriched in Na-N-He and depleted in C and O formed out of gas polluted by the ejecta of stars massive enough to host hot-hydrogen burning.  The nature of the polluters has remained debated, however.  Candidate polluters include asymptotic giant branch stars \citep[mass $\simeq 4-8\,\Ms$, ][]{dant02,derc08,dant16}, fast-rotating massive stars \citep[mass $> 20\,\Ms$, ][]{pra06,dec07}, and super-massive stars \citep[mass $\simeq 10^4\,\Ms$, ][]{den14}.  Given their very different masses, measurements of stellar age spreads in \gcs could help pinpoint the most likely scenario.  Yet, this remains a challenging task, hindered by the old age of \gcs \citep{oli20}.   

To complexify the picture further, the \cm diagram of some younger clusters presents extended main sequence turn-offs, as well as split main sequences (for ages younger than 2\,Gyr and 700\,Myr, respectively).  Observed in the Large Magellanic Cloud clusters and in open clusters of the Galactic disk, such features stem from different stellar rotation velocities \citep{dup17,mar18,cor18}.  

Last but not least, some forming clusters present stellar age distributions with peaks and troughs, suggesting that their \stf activity periodically recedes, to be re-invigorated later on.  Examples are the Orion Nebular Cluster (ONC) in the Galactic disk \citep{bec17}, and the young cluster R136 at the heart of the 30\,Dor HII region in the Large Magellanic Cloud \citep{sch18}.  Interestingly, the age-span between successive peaks is of order 1\,Myr for both the ONC and R136.  It remains unclear whether and how these complex \stf histories relate to the multiple stellar populations of old globular clusters.  
    
To explain the bumpy \stf history of the ONC, \citet{kro18} devise a scenario in which clusters with masses in the range $600-20,000\,\Ms$ form at the converging point of molecular gas inflows.  Newly-born O-stars shut off the inflow and \stf until they get ejected out of the cluster via stellar interactions.  The gas inflow then  resumes and builds the next stellar population.  

We will explore another avenue, one in which the gas density of a cluster-forming clump is regulated by the formation and dispersal of 'failed' cores.
Molecular cores are often considered the precursors of individual stars and small stellar systems \citep[e.g.][and references therein]{and14,bpa20}.  Not all cores are prestellar, however.  Some, possibly the majority, fail to form stars, 'bounce back' and disperse in the clump gas out of which they initially form \citep{tay96,vzs05,cla05,gom07,lad08,bel11a}.  The cycle of their formation and dispersal must therefore impact the mass and density of the clump gas, hence the core and \stf rates.  We consider these 'failed'/'rebounding' /'dispersing' cores as a source of gas replenishment possibly leading to another \stf episode.  

We will investigate whether different regions of the model parameter space can render different \stf histories, from sequences of fairly discrete \stf episodes to smoothly evolving \stf rates.  Such a finding could be a major gateway to understand the large variance observed in the multiple stellar populations of old globular clusters.  Consider the  distributions of stars along their O-Na anti-correlation \citep{car15,cha16} and in their chromosome map \citep{mil17}.  They differ from cluster to cluster, being sometimes more clumpy, sometimes more continuous.  Are the successive \stf episodes of some forming clusters (e.g. the ONC) the low-mass counterpart of the clumpy multiple stellar populations in some old globular clusters?  

For a new \stf episode to rise, the previous one must first recede.  We therefore opt for a model able to predict a decreasing \stf activity in cluster-forming clumps.  We adopt the volume-density driven cluster-formation model of \citet{par13}, which predicts a \sfr declining with time under the assumptions of constant clump mass and size.  This decline is driven by two factors: (i) the decrease with time of the available gas mass due to ongoing star formation in an isolated constant-size clump, and (ii) the corresponding increase of the gas free-fall time \citep[see][for a detailed discussion]{par14}.  Evidence for decreasing \stf histories have been found for the Chamaeleon~I molecular cloud \citep{luh07,bel11b} and, possibly, for the Perseus molecular cloud \citep{hat05} \citep[see also][]{prei12}.  Additionally, the amplitude of the three \stf episodes in the ONC is also a declining function of time \citep{bec17}.  

\citet{par13} assume that the clump gas is directly converted into stars.  Yet, stars form out of molecular cores, which are themselves condensations forming out of the clump gas.  With this in mind, we add two major elements to their model: (1)~the formation of cores constitute an intermediate stage in the conversion of the clump gas into stars, and (2)~some cores - rather than collapsing and forming stars - re-expand into the clump gas reservoir, thereby re-supplying it in gas. 
In this preliminary study, we focus on cluster-forming clumps with uniform density.  

 The outline of the paper is as follow.  In Sec.~\ref{sec:modI}, we introduce a preliminary model, out of  which we  develop a criterion for the rise of a second \stf episode.  Section~\ref{sec:fco} reviews the evidence for failed cores in simulations and observations of \sfing regions.  In Sec.~\ref{sec:modII}, we expand the model of Sec.~\ref{sec:modI} by differentiating between star-forming and failed cores, and we explore the parameter space in Sec.~\ref{sec:gald}.  We further upgrade the model in Sec.~\ref{sec:ipts}, where we relate the dissolution time-scale of the failed cores to the free-fall time of the host clump, using the Orion Nebula Cluster as a test-bed.  In Sec.~\ref{sec:disc}, we discuss how our model \stf histories could be related to \msps in old globular clusters, the limitations of our current model, and how an initial gas density gradient will impact our current results.   We present our conclusions in Sec.~\ref{sec:conc}.

\section{Model Fundamentals} \label{sec:modI}

The model of \citet{par13} predicts that the \sfr of cluster-forming clumps is a decreasing function of time.  This behavior stems from their model clumps having a constant radius and being isolated (i.e. no inflow, no outflow).  Declining \stf rates such as those shown in Fig.~3 in \citet{par14} and Fig.~4 in \citet{par19} define the first \stf episode of our model clumps.  We aim at pinpointing the conditions leading to a second \stf episode, while retaining the assumptions of an isolated clump with a fixed size.  The pivotal point here is: to form stars, a molecular clump must first form starless molecular cores.  Some of these cores are bound, gravitationally unstable, collapse and form a star or a small stellar group (i.e. binary or triple).  Not all cores are prestellar, however.  Others remain unbound and eventually bounce back to the clump gas reservoir out of which they formed \citep[e.g.][see also Sec.~\ref{sec:fco} for a discussion]{tay96,vzs05,cla05,lad08,bel11a}.  
The formation and dispersal of these 'failed' cores equates with removing some gas from the clump gas reservoir (core formation), and restoring it at a later time (core dispersal).  When failed cores are abundant enough, their dispersal  increases the leftover gas mass of a clump, renewing its core and star formation.  To take into account the lifecycle of molecular  cores - especially the non-collapsing ones - will therefore yield \stf histories different from those shown in  \citet{par14} and \citet{par19}.   In what follows, we coin {\it return time} the time-span needed for non-collapsing cores to disperse.  The imprint of this parameter is the time-span between two successive peaks of a \stf history. 

\subsection{Building Bumpy Core Formation Histories} \label{ssec:cofh}
To model the core formation history of a cluster-forming clump, we use Eq.~18 of \citet{par13}, substituting the \cofr per unit volume, $d\rhoco(t)/dt$, for the \sfr per unit volume $d\rhost(t)/dt$, and the \cofe per free-fall time, $\effco$ for the \sfe per free-fall time.  We obtain 

\begin{equation}
\frac{d\rhoco(t)}{dt} = \frac{\effco}{\tff(t)} \rhog(t) = \sqrt{\frac{32G}{3\pi}} \effco \rhog(t)^{3/2}\;.
\label{eq:par13eq18}
\end{equation}

In this equation, $\rhog(t)$ is the volume density of the clump gas at time $t$ after \cof onset, $\tau_{ff}(t)$ is the clump gas \fft and $G$ is the gravitational constant.  For a clear understanding of how the gas density impacts our model, we consider clumps with a uniform density.  The evolution of clumps with an initial gas density gradient will be investigated in a future paper.

Equation \ref{eq:par13eq18} shows that the \cofr per unit volume at a given time $t$ depends on the corresponding gas density, hence on the gas mass inside a constant-radius clump.  The gas mass equates with the clump total mass minus the mass in cores and stars.  At first, ongoing core formation depletes the gas reservoir, thereby decreasing the \cof rate.  The higher the initial density of the clump gas (i.e. the shorter its initial free-fall time $\tffi$) and/or the greater the \cofe per \fft $\effco$, the higher the initial \cof rate, but also the faster the gas density and \cofr dwindle.  We note that a gas density gradient would hasten the system evolution as well \citep[see Fig.~4 in][]{par19}.  

For a second \cof episode to occur, the \cofr must stop declining and start rising.  Eq.~\ref{eq:par13eq18} shows that this happens if the clump gas reservoir is re-supplied in gas.  In our model, this gas comes from cores formed at an earlier time, which now get dispersed following their failure to form stars.  Consider cores formed at time $t\gtrsim0$ (shortly after core formation onset).  After a return time $t_{return}$, a fraction $f_{return}$ of the mass in cores returns to the clump gas.  If the gas supply from dispersing cores overcomes the gas depletion due to ongoing core formation, the clump gas mass and density are on the rise.  The return fraction $\fret$ is obviously an important parameter.  Here we can consider two extreme scenarios.  If cores consist exclusively of non-collapsing ones, the return fraction is unity, $\fret=1$.  If only pre-stellar cores form, then the return fraction is their envelope mass fraction and $\fret \simeq 2/3$ \citep[i.e. the \sfe of a single core is $\simeq 1/3$; ][]{mat00,alv07}.  We therefore adopt an intermediate value $\fret = 0.85$, and we keep it constant.  For now, we assume that \sfing cores release their envelope and failed cores disperse on the same time-scale $\tret$.  We will lift this hypothesis in Sec.~\ref{sec:modII}.  Armed with such a model, we can already anticipate the possibility of generating not just two, but multiple \stf episodes.  Indeed, if the second \cof episode is significant enough, it will feed the emergence of a third one.

Having detailed how our preliminary model unfolds, we now consider a spherical clump of mass $m_{clump}$ and radius $r_{clump}$, and we infer the evolution with time of its gas density and \cof rate.  The initial \fft of the clump gas is
\begin{equation}
\tffi = \sqrt{ \frac{3\pi}{32G\rho_0} }
\label{eq:tff}
\end{equation}
with $\rho_0$ the clump volume density, equivalently the initial gas volume density:
\begin{equation}
\rho_0 = \rho_{gas}(t=0) = \frac{m_{clump}}{\frac{4}{3}\pi r_{clump}^3}\;.
\end{equation}
 
\begin{figure}
\begin{center}
\epsscale{1.10}  \plotone{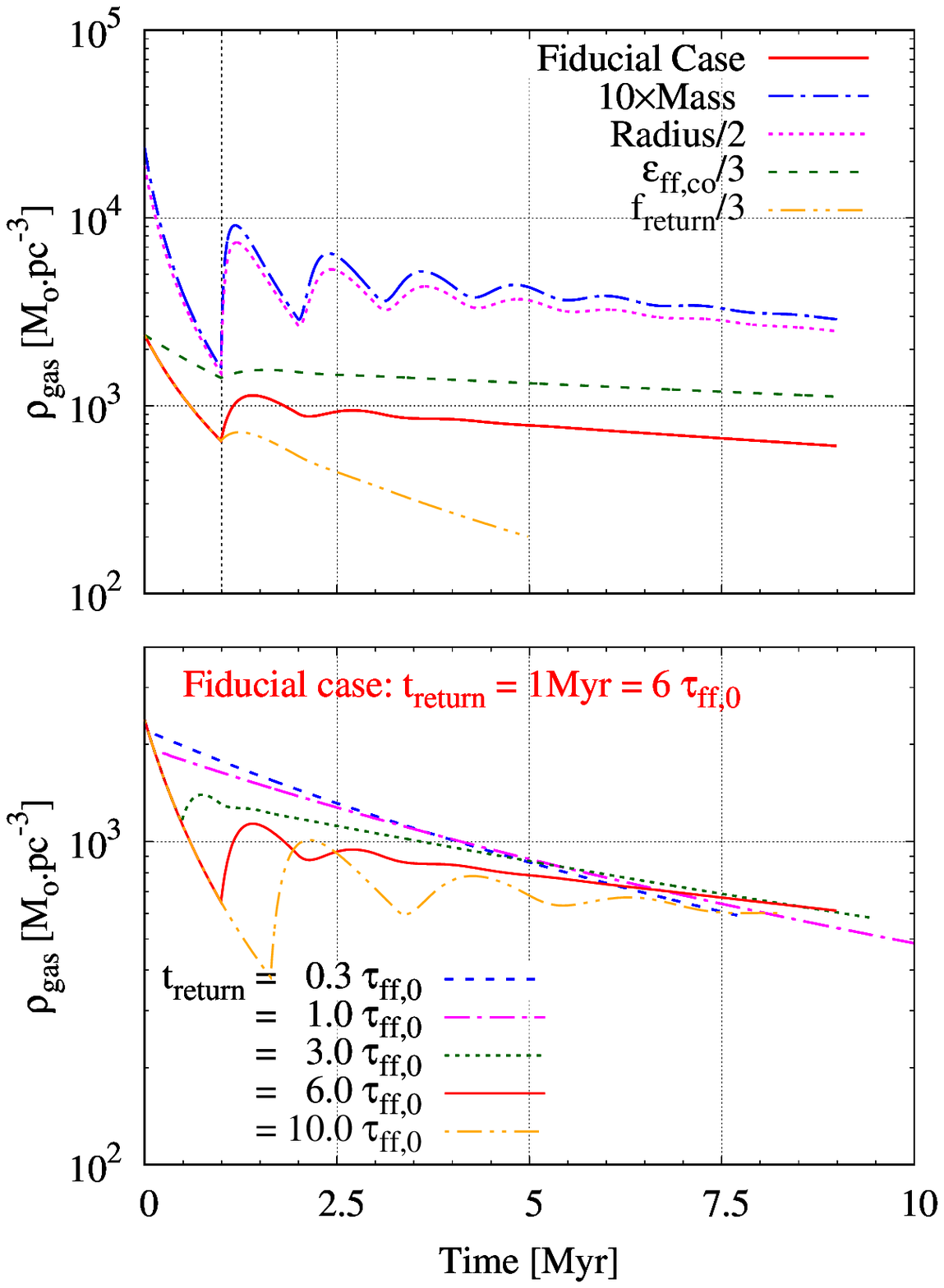}
\caption{Evolution with time of the clump gas density.  Top panel: Results for the fiducial model ($m_{clump}=10^4\,\Ms$, $r_{clump}=1\,$pc, $\effco=0.30$, $\tret = 1$\,Myr and $\fret=0.85$: solid red line), for a ten-times more massive clump ($m_{clump}=10^5\,\Ms$, dash-dotted blue line), for a twice as compact clump ($r_{clump}=0.5$\,pc: dotted magenta line), for a three-times lower \cofe per \fft ($\effco=0.1$: dashed green line), and for a $\sim3$-times lower return fraction of the mass in cores ($\fret=0.3$: dash-dotted orange line).  Bottom panel: Results when varying the return time of the core mass: $\tret/\tffi = 0.3, 1.0, 3.0, 6.0, 10.0$ (see key), all other parameters being identical to the top panel.  The fiducial model (red line) is the same as in the top panel  and corresponds to $\tret/\tffi = 6.0$.  Note the different extents of the $y$-axis in both panels.}
\label{fig:rhog}
\end{center} 
\end{figure}

The top panel of Fig.~\ref{fig:rhog} shows as the solid red line the evolution of the gas density of a clump of mass $m_{clump}=10^4\,\Ms$, radius $r_{clump}=1\,$pc and \cofe per \fft $\effco=0.30$.  The return time and the return fraction of the core gas are set to $\tret = 1$\,Myr and $\fret=0.85$.  
The return time is shown as the vertical dotted black line.  These parameter values define the fiducial model of Fig.~\ref{fig:rhog}.  
As long as $t < \tret$, the evolution of the gas density is as predicted by \citet[][their Eqs~17-18, written for cores rather than for stars]{par13}.  It steadily decreases due to the formation of new cores
\begin{equation}
\frac{d\rhog(t)}{dt} = -\frac{d\rhoco(t)}{dt}\;.
\label{eq:covg}
\end{equation}
At $t\gtrsim\tret$, the gas density increases.  This shows that the dispersal of cores returns enough mass to the clump gas to compensate for the gas used by ongoing core formation.  The evolution of $\rhog(t)$ already tells us that the \cof history will present two well distinct peaks followed by a third weaker one and a long trail of decreasing \stf activity.  

How does modifying the fiducial model affect the time evolution of the gas density?  We consider the following modifications (one at a time): a 10-times more massive clump ($m_{clump}=10^5\,\Ms$), a twice as small clump radius ($r_{clump}=0.5$pc), a three-times smaller \cofe per \fft ($\effco = 0.1$) and an (unphysically) small return fraction of the core mass ($\fret=0.3$).  Results are presented along with the fiducial model in the top panel of Fig.~\ref{fig:rhog}.

Models with a reduced return fraction $\fret$ (orange line) or a reduced \cofe per \fft $\effco$ (green line) fail to replenish significantly the clump gas reservoir at $t=t_{return}$, either because of too great a mass fraction remaining locked inside cores (small $f_{return}$), or because the gas density has not decreased enough to establish a strong gas density contrast between $t = 0$ and $t = \tret$ (small $\effco$).   
 
In contrast, a higher initial gas density (obtained through either a larger clump mass $m_{clump}$ or a smaller clump radius $r_{clump}$) promotes the formation of successive \stf episodes.  This is because the mass fraction of a clump  which gets turned into cores depends on time in units of the free-fall time, $t/\tffi$.  The physical return time of cores being fixed ($\tret =1$\,Myr), the core return time $\tret/\tffi$ is longer for denser clumps.  This builds a stronger contrast between the gas densities at $t=0$ and $t=\tret$ since the clump gas density is given a longer time-span $\tret/\tffi$ to decrease.  In turn, this allows the core gas released at $t\gtrsim \tret$ to make a deeper impact.     

\begin{figure}
\begin{center}
\epsscale{1.10}  \plotone{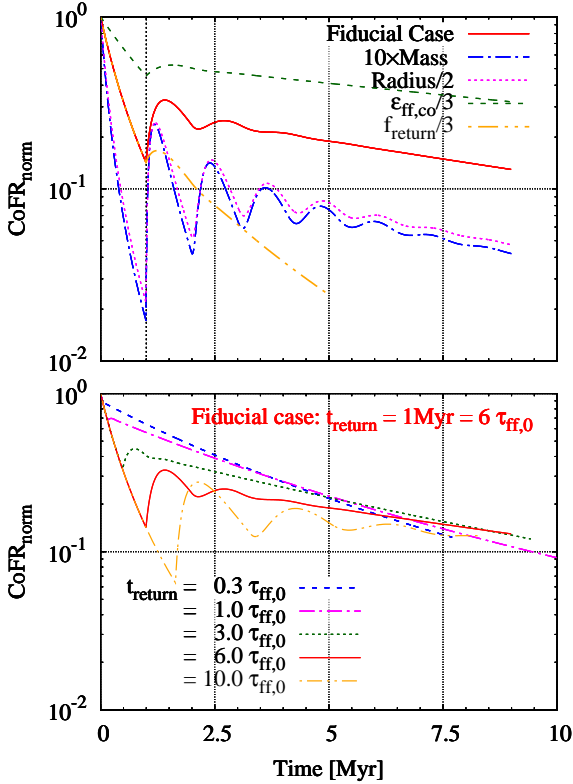}
\caption{Same as Fig.~\ref{fig:cofr} but for the \cof rates normalized to their initial value}
\label{fig:cofr}
\end{center} 
\end{figure}   

The importance of the return time of failed cores is highlighted in the bottom panel of Fig.~\ref{fig:rhog}, where the fiducial model ($\tffi=0.17$\,Myr and $\tret = 1$\,Myr $\equiv 6\,\tffi$) is compared to models with $\tret/\tffi = 0.3, 1.0, 3.0, 10.0$.   Clearly, longer return times $\tret/\tffi$ stimulate more \cof episodes.  

Figure~\ref{fig:cofr} shows the corresponding \cof rates normalized to their initial value (same model parameters and color-coding  as in Fig.~\ref{fig:rhog}).  The variations of the \cofr are wider than those of the gas density, since the \cofr scales as the gas density to the power of 1.5 (Eq.~\ref{eq:par13eq18}).  
Fig.~\ref{fig:cofr} highlights the ability of our model to create very diverse \cof histories.  This provides us with a hand to understand the strong variance observed in cluster multiple stellar populations, an aspect we shall discuss in Sec.~\ref{sec:disc}.

The top panel of Fig.~\ref{fig:cofr} may prompt us to conclude that dense and massive clumps are more likely to generate multiple \cof episodes.  This conclusion is erroneous, however, and we will return to this important aspect also in Sec.~\ref{sec:disc}.  Here, we stress that the model parameters shaping a \cof history are the {\it dimensionless} parameters, namely, the  \cofe per \fft $\effco$, the return fraction $\fret$ and return time $\tret/\tffi$ of the core mass.  We  now develop a tool to quantify their impact.  

\subsection{Criterion for a second core formation episode} \label{ssec:epi2}
The rise of a second \cof episode requires the gas density, in the aftermath of its steady decrease since \cof onset, to make an upturn and to start increasing at $t=\tret$.  This is only doable if non-collapsing cores return to the clump gas a mass greater than that "pumped" to form new cores.  Building on Eq.~\ref{eq:par13eq18}, the following equation  expresses the clump gas density at $t=\tret+dt$, $\rhog(\tret+dt)$, as a function of its counterpart at $t=\tret$,  $\rhog(\tret)$:

\begin{equation}
\begin{aligned}
\rhog(\tret+dt) & = \rhog(\tret) \\ & - \sqrt{\frac{32G}{3\pi}} \effco \rhog(\tret)^{3/2} \cdot dt \\
& + \fret \sqrt{\frac{32G}{3\pi}} \effco \rho_0^{3/2} dt\;. 
\label{eq:epi2cri}
\end{aligned}
\end{equation}

The second term on the right-hand side corresponds to the clump gas used by ongoing core formation.  The third term accounts for the cores formed over the time-span $[0,{\rm d}t]$ that return to the clump gas over the time-interval $[\tret,\tret+{\rm d}t]$.  This third term depends on the clump initial gas density $\rho_0$, for this is the gas density at $t=0$ which dictates the formation rate of the cores being dispersed at $t \gtrsim t_{return}$.  Following a few developments, we obtain the slope of the evolution with time of the gas density at $t=\tret$:

\begin{equation}
\begin{aligned}
\left[\frac{d\rhog(t)}{dt}\right]_{t=\tret}    
 & = \frac{\rhog(\tret+dt) - \rhog(\tret)}{dt}   \\
 & = \frac{CoFR(t=0)}{V}  
 \cdot S(\effco, \fret, \tret/\tffi) 
\label{eq:epi2cri}
\end{aligned}
\end{equation}
with 
\begin{equation}
 S(\effco, \fret, \tret/\tffi) 
 = \left[ \fret - \left( 1 + \frac{1}{2} \effco \frac{\tret}{\tffi} \right)^{-3} \right]\;. 
\label{eq:S}
\end{equation}

In Eq.~\ref{eq:epi2cri}, $V$ is the clump volume and $CoFR(t=0)$ is the clump initial \cof rate:

\begin{equation}
\frac{CoFR(t=0)}{V}=\left[ \frac{d\rhoco(t)}{dt} \right]_{t=0} = \sqrt{\frac{32G}{3\pi}} \effco \rho_0^{3/2}\;.
\label{eq:CoFRi}
\end{equation}

The function $S(\effco, \fret, \tret/\tffi)$, given by Eq.\ref{eq:S}, determines whether the slope of $\rho_{gas}(t)$ turns positive at $t=\tret$, or not.  For instance, if the cores return no gas ($\fret=0$), then $S<0$ and $\left[\frac{d\rhog(t)}{dt}\right]_{t=\tret} < 0$.  That is, the gas density keeps decreasing and there is no second \cof episode.  Conversely, if $S(\effco, \fret, \tret/\tffi)>0$, $t=\tret$ corresponds to the onset of a second \cof episode.  

\begin{figure}
\begin{center}
\epsscale{1.10}  \plotone{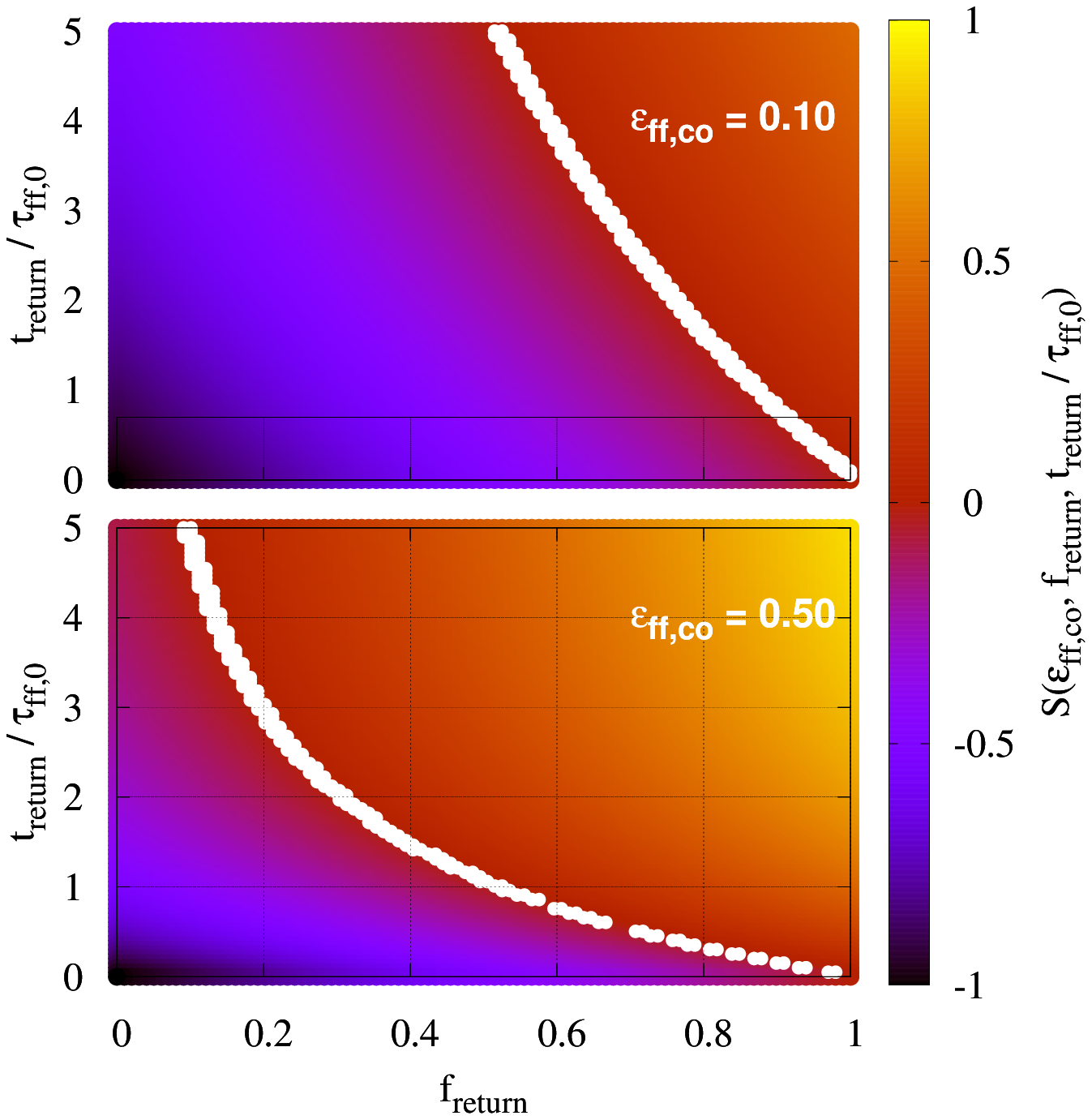}
\caption{The  $S$-function in dependence of the \cofe per \fft $\effco$, and of the return fraction $\fret$ and return time $\tret/\tffi$ of the mass in cores (Eq.~\ref{eq:S}).  It is color-coded according to the right-hand-side bar.  When $S>0$ (red-to-yellow area), the clump gas density at $t=\tret$ increases, possibly yielding a second \cof episode.  When $S<0$ (red-to-black area), the clump gas density at $t=\tret$ keeps decreasing, making impossible a second \cof episode.  The white symbols depict the locus of points for which $S=0$.}
\label{fig:S}
\end{center} 
\end{figure}   
  
Figure~\ref{fig:S} maps $S(\fret, \tret/\tffi)$ for $\effco=0.1$ and $\effco=0.5$.  Its value is color-coded according to the right-hand-side bar, and the white symbols part the region for which $S>0$ from that with $S<0$.  When $S>0$ (red-to-yellow area), the clump gas gains, at $t \gtrsim \tret$, more mass from failed cores formed at $t\gtrsim 0$ than it loses to ongoing core formation.  When $S<0$ (red-to-black area), the formation of new cores consumes more clump gas than what failed cores provide. 
The higher the \cofe per free-fall time, the wider the area of the $(\fret, \tret/\tffi)$ parameter space for which $S>0$.   For a given \cofe per free-fall time, a long return time of the core gas $ \tret / \tffi$ and a large return mass fraction $\fret$ favor a gas density increase at $t=t_{return}$ and, therefore, an increase of the \cof rate.  Such conclusions have already  been suggested by Fig.~\ref{fig:rhog}.    

\noindent We note that since, at \cof onset, 
\begin{equation}
\left[\frac{d\rhog(t)}{dt}\right]_{t=0} = - \left[\frac{d\rhoco(t)}{dt}\right]_{t=0}=-\frac{CoFR(t=0)}{V}\,,
\end{equation}
Eq.~6 can be rewritten as:
\begin{equation}
\begin{aligned}
\left[\frac{d\rhog(t)}{dt}\right]_{t=\tret} = &
 - \left[\frac{d\rhog(t)}{dt}\right]_{t=0} \\
& \times  S(\effco, \fret, \tret/\tffi)\;.
\label{eq:drhocomp}
\end{aligned}
\end{equation}

As can be seen, $S$ compares how fast the gas density varies at $t=\tret$ and $t=0$.  Since the upper limit on $S$ is $1$ (when $\fret=1$ and $\tret/\tffi \rightarrow \infty$), Eq.~\ref{eq:drhocomp} expresses that, in absolute value, gas density variations at $t \gtrsim \tret$ cannot be faster than those at $t \gtrsim 0$.  

The model parameters promoting a second core formation episode are a high \cofe per free-fall time $\effco$, a large mass fraction $\fret$ of core gas not ending into stars, and a long return time of the core gas to the clump gas $\tret/\tffi$.    
A second core formation episode can be compared to a bouncing ball.  A ball bounces back if (1)~ it is dropped from a high height, and (2)~its encounter with the ground is as elastic as possible, such that it does not lose much of the kinetic energy accumulated through its initial fall.  The height equivalent in our model is $\effco \tret / \tffi$, which determines  which fraction of the clump mass has been turned into cores at $t=\tret$: the higher $\effco$, the longer $\tret/\tffi$, the greater the initial gas mass fraction turned into cores, and the more the gas density has dropped compared to its initial value.  As for our $\fret$ return fraction, its bouncing ball counterpart is the near-conservation of the ball kinetic energy: the higher $\fret$, the higher the core gas mass released at $t=\tret$, and the higher the gas density and \cofr - like the ball - bounce back.
 
The high \cof efficiencies per \fft required for multiple \cof episodes (e.g. $\effco = 0.30$) are to be contrasted with the \stf efficiencies  per free-fall time measured in the Galactic disk.  The latter are of order one per cent \citep{kt07,pok21}.  Given that individual star-forming cores have a \sfe of $\simeq 1/3$ \citep{mat00,alv07}, the formation efficiency per \fft of  prestellar cores should not exceed a few per cent.  Multiple \cof episodes therefore require non-collapsing cores to dominate the core population.  This is also indicated by our high return fraction ($f_{return}=0.85$).  We now review the evidence for significant populations of failed cores in molecular gas clouds, as found by observations and simulations.

\section{Evidence for failed cores in molecular clouds} \label{sec:fco}
\citet{bel11a} map the nearby molecular cloud Chamaeleon~I in dust continuum emission at 870$\micron$.  They find that 17\% of its starless cores have a mass higher than the critical Bonnor-Ebert mass\footnote{The Bonnor-Ebert mass is the critical mass beyond which isothermal cores cannot be in hydrostatic equilibrium and collapse \citep{bon56,tan14}.}.  That is, only a minority of the Chamaeleon~I cores are gravitationally unstable, and they tend to be the most massive ones.  Additionally, the mass function of the starless cores is over-populated at its low-mass end, while being consistent with the Salpeter (1955) slope at the high-mass end.  \citet{bel11a} therefore conclude that most starless cores in Chamaeleon~I are not prestellar.  \citet{bel11b} further suggest that some of these low-mass gravitationally-stable cores may never reach their critical mass and "get dispersed" in the future.  Their  interpretation is supported by a number of numerical  simulations.

According to \citet{cla05}, most of the clumps (their terminology for cores) produced by supersonic turbulence are  unbound.  In one of their simulations, unbound clumps outnumber bound ones by two orders of magnitude.  Unbound clumps/cores tend to be the low-mass ones, while prestellar cores are the most massive ones (since they have to reach the Jeans mass in the first place).  A consequence of this is the absence of a one-to-one mapping from cores to stars and, therefore, from the core mass function to the stellar initial mass function \citep{kle01,cla05}.

The nearby Pipe molecular cloud presents us with yet another population of failed cores.
159 cores with a mean volume density $\simeq 7\cdot 10^3\,\cc$ are detected by \citet{lad08}.  A comparison of their escape velocity with their - mostly thermal - gas velocity dispersion show most of them to be unbound \citep[][their fig.~4]{lad08}.  The transition between the bound and unbound regimes occurs at a core mass of $2-3\,\Ms$, with 140 cores being less massive than $3\,\Ms$.  Unless they experience a decrease of their internal pressure, or an increase of their external pressure (or a combination of both), such cores are unlikely to be prestellar.  \citet{lad08} stress that unbound cores should persist for one or more sound crossing-times before they disperse in their surroundings \citep[see also ][]{ket06}.  The radius of the unbound cores is of order 0.075\,pc \citep[][their Fig.~6]{lad08} and the sound speed in an isothermal gas at a temperature of $10$K is $c_s = 0.2\,km\cdot s^{-1}$.  The sound crossing time of the unbound cores is therefore $\tau_{core, cross} = 2 \cdot 0.075\,{\rm pc}/(0.2\,{\rm pc} \cdot {\rm Myr}^{-1})  \simeq 0.75$\,Myr.  Unbound cores in the Pipe cloud should thus persist as coherent entities for about 1\,Myr, possibly more \citep{lad08}.     

Based on MHD simulations, \citet{vzs05} too find that some cores/clumps form and dissolve later on.  Considering a turbulent isothermal molecular cloud with a mean number density $n_0 = 500\,\cc$, they infer a life-expectancy of 0.5 to 3.0\,Myr for its non-collapsing cores, similar to what \citet{lad08} estimate for the Pipe cloud cores.  Note that the simulated cloud and the Pipe cloud have mean densities within an order of magnitude of each other \citep[the mean density of the Pipe cloud is $\simeq 100\,\cc$, ][]{kai14} .    

The life-expectancy of failed cores equates with the return time $t_{return}$ of our preliminary model and, therefore, with the periodicity with which a clump unfolds its successive \stf episodes.  Both \citet{vzs05} and \citet{lad08} estimate this life-expectancy to be of order 1\,Myr.  Interestingly, this is similar to the time-span separating the three stellar populations of  the ONC \citep[][their Fig.~6]{bec17}, and Fig.~5 of \citet{sch18} reveals a similar periodicity for the R136 cluster in 30~Dor.  We shall come back to this in Sec.~\ref{sec:ipts}.

\section{Failed- versus Star-Forming Cores} \label{sec:modII}
The preliminary model of Section \ref{sec:modI} assumes that \sfing and failed cores release their gas on the same time-span $t_{return}$.  Yet, both types of core evolve on different time-scales.  First, \citet{vzs05} stress that the re-expansion of failed cores is slower than their initial compression, as a result of self-gravity.  Second, their MHD simulations show that failed cores are often less dense than prestellar ones, implying a longer free-fall time.  Therefore, for at least two reasons (self-gravity and longer free-fall time), failed cores evolve more slowly than star-forming ones.  This will help us meet one of the criterions unveiled in Sec.~\ref{sec:modI}, namely, that failed cores should have long return times $\tret/\tffi$ to stimulate a second \cof episode.  In this section, we assign different time-scales and different return fractions to failed and \sfing cores.   In the case of \sfing cores, their envelope is the mass they return to their surroundings \citep{mat00}.   

We assume that collapsing/\sfing cores have a mean density two orders of magnitude higher than that of their host clump.  Their \fft $\tau_{ff, colco}$ is thus ten times shorter than the clump's: $\tau_{ff,colco} = 0.1\tffi$.  Their lifetime ranges from one to ten times their free-fall time, depending on the magnetic field \citep[e.g.][and references therein]{vzs05,gmd07,wt07}.   
We therefore assume that the time-span between the formation of prestellar cores and the release of their envelope is 3 times their free-fall time.  This defines the return time of their envelope to the clump gas as 
\begin{equation}
t_{ret,colco} = 3\tau_{ff,colco} = 0.3 \tffi\;.
\label{eq:tretcolco}
\end{equation}
Models yielding multiple core and \stf episodes hinge on failed cores outnumbering \sfing ones (see Fig.~\ref{fig:S} and below).   The exact values of the density contrast between \sfing cores and their host clump and of $t_{ret,colco}/\tffi$ are thus minor issues.

As for the return time of the failed cores, namely, the time-span on which they re-expand into their surroundings, we fix it to 1\,Myr (see Sec.~\ref{sec:fco}): 
\begin{equation}
t_{ret,fco} = 1\,{\rm Myr}\;.
\label{eq:tretf}
\end{equation}
We will lift this assumption in Sec.~\ref{sec:ipts}, a single physical return time $t_{ret,fco}=1$\,Myr for failed cores being  an oversimplification, especially when considering very different host clump densities $\rho_0$.   

Let us now quantify the core mass fractions (i.e. return fractions) associated to the return times defined above.  We introduce the \sfe per \fft $\effst$ in our model and derive the return fractions as a function of $\effco$ and $\effst$.
The mass $dm_{co}(t)$ in cores formed over a time interval $[t, t+dt]$ consists of three components: (1,2) stars and envelopes formed and released, respectively, by \sfing cores, and (3) failed cores.  The following set of equations relates the mass fraction of failed cores, $F_{fco}$, to the mass fractions of envelopes and stars from \sfing cores, $F_{env}$ and $F_{*}$, respectively:
\begin{equation}
1 = F_{fco} + F_{*} + F_{env}\;.
\label{eq:mf1}
\end{equation}
\begin{equation}
F_{env} = 2F_{*}\;.
\label{eq:mf2}
\end{equation}
\begin{equation}
\frac{\effst}{\effco} = F_{*} 
\label{eq:mf3}
\end{equation}
For \sfing cores, we assume that the mass of their  envelope is twice the mass of their stars (Eq.~\ref{eq:mf2}), i.e. the \sfe of a \sfing core is one-third \citep{mat00,alv07}.  
Eq.~\ref{eq:mf3} quantifies which fraction $F_{*}$ of the total mass in cores eventually ends into stars.  We define it as the ratio of the \sfe per \fft $\effst$ of the host clump to its \cofe per free-fall time $\effco$.  Note that $\effco$ accounts for all cores, both failed and \sfing ones.  For an assumed pair ($\effco$, $\effst$), the above system of equations yields the respective contributions of stars, \sfing core envelopes, and failed cores to the core mass $dm_{co}(t)$ formed over the time-span $[t,t+dt]$.  Note that in Sec.~\ref{ssec:epi2}, $\effco$ and $\fret$ were independent of each other.  With the above set of equations and a fixed $\effst$, $\effco$ and $F_{fco}$ are now tied to each other, an increase of $\effco$ implying an increase of $F_{fco}$. 

\begin{table}
\begin{center}
\caption{Mass fractions of the stars ($F_*$) and envelopes ($F_{env}$) from \sfing cores, and of the failed cores ($F_{fco}$), for given $\effco$ and $\effst$. } 
\begin{tabular}{c c c c } \tableline 
~$\effco$~ & ~$F_{*}$~ & ~$F_{env}$~ & ~$F_{fco}$~  \\ \tableline
\multicolumn{4}{c}{$\effst=0.01$} \\ \tableline
0.030      & 0.33      &  0.67      &  0.00     \\
0.100      & 0.10      &  0.20      &  0.70     \\ 
0.300      & 0.03      &  0.07      &  0.90     \\ 
0.500      & 0.02      &  0.04      &  0.94     \\ \tableline
\multicolumn{4}{c}{$\effst=0.026$} \\ \tableline
0.078      & 0.333      &  0.667      &  0.000     \\
0.300      & 0.087      &  0.173      &  0.740     \\ 
0.500     & 0.052      &  0.104      &  0.844     \\ \tableline
\end{tabular}
\label{tab:t1}
\end{center}
\end{table}     

Table~\ref{tab:t1} provides examples for various \cof efficiencies per free-fall time $\effco$ and for  $\effst=0.01$ (\citet{kt07}; see also Fig.~4 in \citet{pp20}) and $\effst=0.026$ \citep{pok21}.  
When $\effst=0.01$ and $\effco = 0.03$ (or when $\effst=0.026$ and $\effco = 0.078$), failed cores make no contribution to the core mass budget (since $F_{*} = \effst / \effco = 1/3$ implies $F_{env} = 2/3$ and $F_{fco}=0$).  In the absence of failed cores, the only gas returned to the \icm is that of \sfing core envelopes, which, per clump free-fall time, amounts to a few per cent of the clump gas (i.e. twice the \sfe per free-fall time).  This is an extremely small fraction compared to the $\gtrsim 90$\,\% of the clump mass which, per free-fall time, remain in its gas reservoir.  Unsurprisingly, a model with no failed cores fails to generate multiple peaks in the \cof history (see below).  As the \cofe per \fft $\effco$ increases - while retaining the same \sfe per \fft $\effst$ - the contribution of \sfing cores $F_{env} + F_{*}$ dwindles, while that of failed cores $F_{fco}$ rises (see Table \ref{tab:t1}). 

\begin{figure}
\begin{center}
\epsscale{1.10}  \plotone{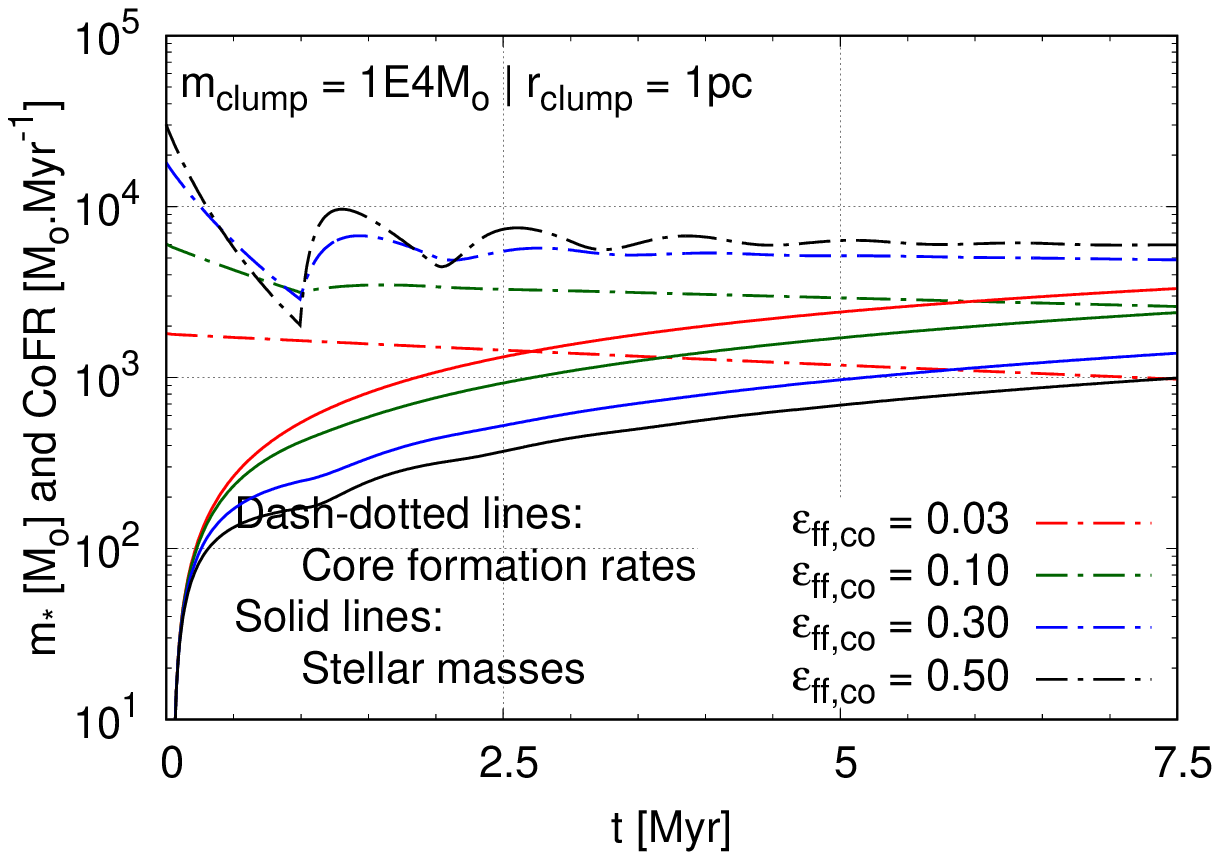}
\caption{Evolution with time of the \cofr (dash-dotted lines) and stellar mass (solid lines) of a clump with an initial gas mass $m_{clump}=10^4\,\Ms$, a radius $r_{clump}=1$\,pc and a \sfe per \fft $\effst=0.01$.  With 
higher \cof efficiencies per \fft $\effco$, failed cores contribute more to the core mass budget (see Table \ref{tab:t1}), and multiple peaks get carved in the \cof history}
\label{fig:2ts}
\end{center} 
\end{figure}   

The four cases of Table~\ref{tab:t1} with $\effst=0.01$ are illustrated in Fig.~\ref{fig:2ts}.  The mass and radius of the model clump are $m_{clump}=10^4\,\Ms$ and $r_{clump}=1$\,pc, respectively, yielding an initial \fft $\tffi=0.167$ Myr for the clump gas.  The return times of the core gas to the \icm are as given by Eqs \ref{eq:tretcolco} ($t_{ret,colco}=0.3\tffi=0.05$\,Myr) and \ref{eq:tretf} ($t_{ret,fco}=1\,{\rm Myr}=6\tffi$).  Dash-dotted and solid lines show the time evolutions of the \cofr and of the stellar mass, respectively.  Note that stellar masses start rising at $t=t_{ret,colco}$, that is, when the first stars emerge from their gaseous cocoon.       

A higher \cofe per free-fall time yields a higher \cofr initially, but also a faster decrease of this one since the greater gas fraction trapped in cores reduces the clump gas mass accordingly.  An increasingly steeper gas density gradient inside clumps would produce a similar effect \citep[Fig.~4 in][]{par19}.  At $t \gtrsim t_{ret,fco} =1$\,Myr, i.e. after one return time of the failed cores, the \cofr may make an upturn, a rise borne by the gas supplied by the failed cores formed at $t \gtrsim 0$\,Myr.  The reasons as to why larger $\effco$ drive more \stf episodes is two-fold: (1)~as in Sec.~\ref{ssec:epi2}, a larger $\effco$ implies a greater decrease of the gas density after a given time-span $t/\tffi$ (that is, the bouncing ball is dropped from a greater height), and (2)~larger $\effco$ with a fixed \stf efficiency per \fft $\effst$ implies greater return fractions from failed cores (see Table~\ref{tab:t1}; that is, the ball retains more of its kinetic energy while bouncing).  These two effects can be visualized in Fig.~\ref{fig:S} as (1)~a widening of the region for which $S>0$ (i.e. the white symbols and the red-to-yellow area are brought downwards for increasing $\effco$), and (2)~ an increase of $\fret$ (i.e. moving to the right of the horizontal axis). 

That a higher \cofe per \fft is more likely to generate successive \cof episodes comes at a cost, however:  higher $\effco$ yield lower stellar masses at any given time $t$ (see solid lines in Fig.~\ref{fig:2ts}).  This reduced \sfe reflects the larger mass in failed cores produced for higher $\effco$, and the correspondingly lower gas masses available for further core formation.  This effect is not tremendous as increasing $\effco$ by a factor of $17$ (i.e. $0.50/0.03$) reduces stellar masses and \stf efficiencies by a factor $\simeq 3$.

\section{Peak number, gas density and core formation efficiency} \label{sec:gald}
The color-magnitude diagram of the ONC presents 3 well-defined pre-main sequences, possibly suggesting three stellar populations with an age separation of about 1\,Myr \citep{bec17}.  That is, the ONC would have experienced three successive core and \stf episodes.
Building on the model refined in the previous section, we now explore which \cofe per \fft $\effco$ and which initial gas density $\rho_0$ are required to carve peaks in the \stf history of a clump with an imposed periodicity of 1\,Myr.  The results will be used in the next section to constrain the properties of the gaseous progenitor of the ONC based on its observed \stf history.  

\begin{figure}
\begin{center}
\epsscale{1.10}  \plotone{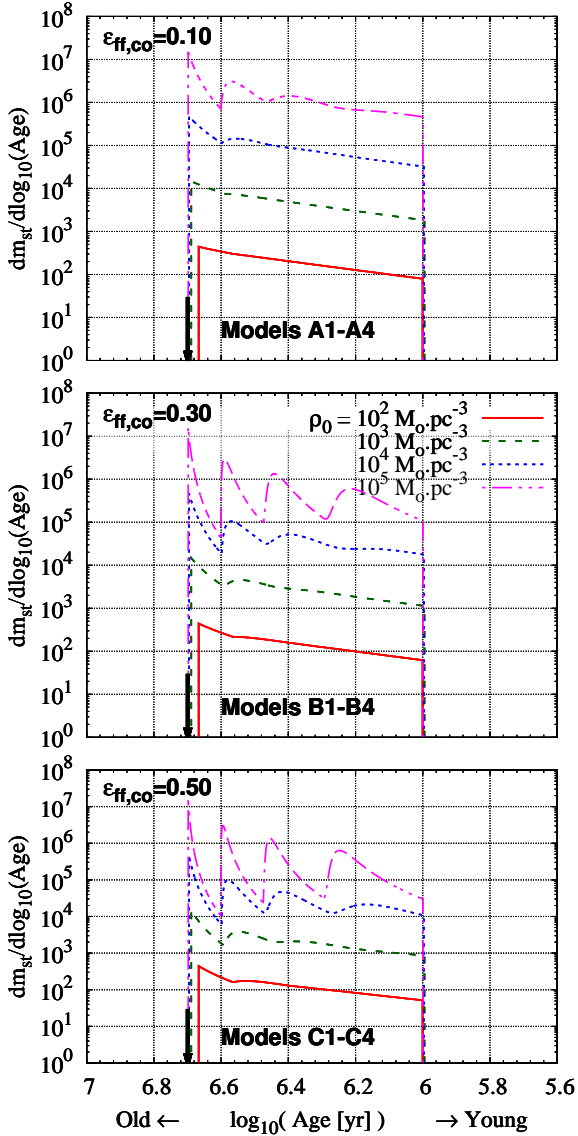}
\caption{Logarithmic stellar age distributions for 3 \cof efficiencies per \fft (from top to bottom panel: $\effco =0.1, 0.3, 0.5$), and 4 initial gas volume densities $\rho_0$ (see key in middle panel).  The adopted densities are representative of  cluster-forming regions of the Galactic disk.  Model parameters are provided in Table \ref{tab:t2}.  }
\label{fig:1myr}
\end{center} 
\end{figure}   

We test three \cof efficiencies per free-fall time, namely, $\effco =0.1, 0.3, 0.5$.  The \sfe per \fft is fixed, and we adopt $\effst=0.01$.  The corresponding core mass fractions converted into stars ($F_*$) and released to the clump gas as failed cores ($F_{fco}$) are given in Table \ref{tab:t1}.  As for the clump initial gas density, we test values ranging from $\rho_0=10^2$ to $10^5\,\Msppp$ in logarithmic steps of 1.  The lower bound is of order the current mean density of the ONC, namely, $4500\,\Ms$ of stars and gas enclosed within a radius of 2\,pc \citep[][]{hh98}.  The upper bound corresponds to the density of the "ultra-high density material" mapped at pc-scale by \citet{hac20} in the active \sfing cloud OMC-1, located in the vicinity of the ONC.  Densities in the range $10^4{\rm -}10^5\,\Msppp$ are also those characterizing the central regions of starburst clusters in the Galactic disk \citep[e.g. Westerlund~1 and NGC\,3603; see top panel of Fig.~10 in][]{par14}.   
To vary the clump density, we vary its mass $m_{clump}$ while keeping its radius $r_{clump}$ constant.  We assume $r_{clump}=2$\,pc, which is approximately the current ONC radius  \citep{hh98,drio17}.  The corresponding clump masses are given in Table \ref{tab:t2}.  All models are assigned a \stf duration of 4\,Myr.  We note that stellar feedback is not taken into account, and that stellar-feedback driven gas expulsion may terminate \stf at an earlier time than assumed here  \citep[e.g.][]{rah17,kro18}.  We impose a physical return time $t_{ret,fco} = 1$\, Myr for the failed cores, as dictated by the time-span separating the ONC three stellar populations \citep[see Fig.~6 in ][]{bec17}.  Combining the clump density with the physical return time of the failed cores provides their return time in units of the clump \fft $t_{ret,fco}/\tffi$ (see Table~\ref{tab:t2}).  The return time of failed cores will be discussed further in Sec.~\ref{sec:ipts}.  With $t_{ret,fco}=1$\,Myr and a \stf duration of 4\,Myr, we expect our model \stf histories to present from 0 to 4 peaks, depending on the \cofe per \fft $\effco$ and failed core return time $t_{ret,fco}/\tffi$.  

\begin{table}
\begin{center}
\caption{Input parameters $\effco$ and $\rho_0$ for the models of Fig.~\ref{fig:1myr}.  $r_{clump}=2$\,pc, $\effst=0.01$ and $t_{ret,fco}=1$\,Myr are imposed.  $m_{st,4Myr}$ is the stellar mass at $t=4$\,Myr} 
\begin{tabular}{c c c c c c } \tableline 
Model & ~$\effco$~ & ~$\rho_0$~ & $t_{ret,fco}$&~$m_{clump}$~     & ~$m_{st,4Myr}$~  \\
 ID &                  & ~$ [\Msppp]$ & [$\tffi$]& ~$[\Ms]$~     & ~$[\Ms]$~  \\ \tableline
  A1  & 0.10      & $10^2$         & ~1.2             &  $3.3\cdot10^3$     &  $1.3\cdot10^2$     \\
 A2  &      0.10      & $10^3$     &  ~3.9              &   $3.3\cdot10^4$     &  $3.5\cdot10^3$     \\
   A3  &    0.10      & $10^4$     & 12.4               &    $3.3\cdot10^5$     & $6.9\cdot10^4$     \\
  A4  &    0.10      & $10^5$      &  39.1             &    $3.3\cdot10^6$      &  $1.0\cdot10^6$     \\ \hline
  B1  &    0.30      & $10^2$     & ~1.2                 &  $3.3\cdot10^3$     &  $1.1\cdot10^2$     \\
 B2  &      0.30      & $10^3$    & ~3.9                 &   $3.3\cdot10^4$     &  $2.2\cdot10^3$     \\
 B3  &      0.30      & $10^4$    & 12.4                &    $3.3\cdot10^5$     & $3.4\cdot10^4$     \\
  B4  &    0.30      & $10^5$     & 39.1                 &    $3.3\cdot10^6$      &  $0.4\cdot10^6$     \\ \hline
  C1  &     0.50      & $10^2$   & ~1.2                    &  $3.3\cdot10^3$     &  $0.9\cdot10^2$     \\
  C2  &     0.50      & $10^3$   & ~3.9                  &   $3.3\cdot10^4$     &  $1.6\cdot10^3$     \\
  C3  &     0.50      & $10^4$   & 12.4                  &    $3.3\cdot10^5$     & $2.3\cdot10^4$     \\
  C4  &     0.50      & $10^5$   &  39.1               &    $3.3\cdot10^6$      &  $0.3\cdot10^6$     \\ \tableline
\end{tabular}
\label{tab:t2}
\end{center}
\end{table}     

The logarithmic stellar age distributions are presented in Fig.~\ref{fig:1myr}.  The downward black arrows at an age of 5\,Myr indicate the onset of core formation.  Note in that respect the slight shift with the onset of the stellar age distribution, especially noticeable for the lowest-density models (i.e. $\rho_0=100\,\Msppp$; red lines).  This small shift corresponds to the time-delay $t_{ret,colco}=0.3\tffi$ required by star-forming cores to shed their envelope in the intra-cluster medium (Eq.~\ref{eq:tretcolco}).      

Figure~\ref{fig:1myr} reveals that the presence of one peak per Myr for $\geq3$Myr requires a clump density of at least $\rho_0 \simeq 10^3\,\Msppp$ when the \cofe per \fft is $\effco = 0.5$ (dashed green line in bottom panel).  Lower \cof efficiencies per \fft must be "compensated" with higher initial gas densities (e.g. $\rho_0 \simeq 10^5\,\Msppp$ when $\effco = 0.1$, dashed magenta line in top panel).  We note that models with a density similar to that of the present-day ONC (i.e. $\rho_0 \simeq 10^2\,\Msppp$) fail to carve multiple peaks in their \stf history, irrespective of the adopted \cofe per free-fall time (solid red lines in all panels).

How can we disentangle between the models with $\rho_0 = 10^3\,\Msppp$ and $\effco=0.5$ and $\rho_0 = 10^5\,\Msppp$ and $\effco=0.1$ (models C2 and A4 in Table~\ref{tab:t2})?  The initial gas mass $m_{clump}$ and the stellar mass $m_{st}$ at $t=4$\,Myr constitute powerful tools here (see last two columns of Table~\ref{tab:t2}).   The C2 model ($\rho_0 =10^3\,\Msppp$ and $\effco=0.5$) has formed $1.6\cdot10^3\,\Ms$ in stars over 4\,Myr (or $1.4\cdot10^3\,\Ms$ over 3.3\,Myr, which is the ONC \stf duration we will infer in the next section).  This mass agrees much better with the current stellar content of the ONC \citep[$1800\,\Ms$;][]{hh98} than the stellar masses predicted with $\rho_0=10^5\,\Msppp$.  Given the adopted radius $r_{clump}=2$\,pc, an initial gas density $\rho_0=10^5\,\Msppp$ implies an initial gas mass of several million solar masses.  This is 5 orders of magnitude higher than the mass of "ultra-high density material" detected by \citet{hac20} in OMC-1 ($\simeq 30\,\Ms$ at $\simeq 10^5\,\Msppp$).  The comparison of the observed gas and star masses to their model counterparts discards therefore the model A4.  In other words, considering both the shape and normalisation of a \stf history allows one to disentangle whether the most appropriate model combines a high \cofe per \fft to a low-density clump, or the other way round.  

In the case of the ONC, we discard models with $\rho_0 = 10^5\,\Msppp$ not just because of their too large gas and stellar masses, but also because the imposed physical return time $t_{ret,fco}=1\,$Myr equates with $t_{ret,fco}=40\tffi$ ($\tffi = 0.026$\,Myr).  Our model assumptions (e.g. constancy of the clump radius $r_{clump}$) are unlikely to remain valid that long.       

\section{The inter-peak age-span as a density probe} \label{sec:ipts} 

A further observable which helps shed light on the initial gas density of a cluster-forming clump is the age span separating successive peaks of its stellar age distribution.  Since high-density systems evolve faster than low-density ones, we intuitively expect the frequency of the peaks to be higher when the clump density is larger.  Yet, we have so far assumed that the return time of failed cores $t_{ret,fco}$ is  independent of other model parameters.  
We now lift our assumption of a fixed return time $t_{ret,fco}=1\,$Myr for the failed cores, and we relate $t_{ret,fco}$ to  the clump \fft $\tffi$, as already done for $t_{ret,colco}$, the time needed for \sfing cores to shed their envelope  (Eq.~\ref{eq:tretcolco}).

\subsection{The return time of the failed cores} \label{ssec:tretfco}
\citet{vzs05} infer an analytical estimate of how long it takes for a re-expanding core to double its radius, that is, to drop its mean density by almost an order of magnitude, when the external pressure exerted by the surrounding gas is negligible.  We adopt this expansion time as a proxy to the failed core return time $t_{ret,fco}$.  The left panel of Fig.~1 in \citet{vzs05} shows that the return time $t_{ret,fco}$ is in the following range
\begin{equation}
2 \frac{R_{eq}}{c_s}\lesssim t_{ret,fco} \lesssim  6 \frac{R_{eq}}{c_s},
\end{equation}  
with $R_{eq}$ the core equilibrium radius and $c_s$ the core gas sound speed.  The exact value depends on how far from equilibrium the core is following its initial compression. Since: 
\begin{equation}
\frac{R_{eq}}{c_s} = \sqrt{ \frac{\pi}{G \rho_{co}} }
\end{equation} 
for a core density $\rho_{co}$ \citep[e.g.][]{bt87}, we can write\footnote{It is worth noting here that \citet{vzs05} adopt $R_{eq}/c_s$ as their definition of the free-fall time, which is therefore longer than ours by a factor $\sqrt{32/3}$.}:
\begin{equation}
t_{ret,fco} = \top \sqrt{ \frac{\pi}{G \rho_{fco} }} = \top \sqrt{\frac{32}{3}} \sqrt{ \frac{3\pi}{32G \rho_{fco}}}  = 3.3 \top \tau_{ff,fco} 
\label{eq:tretfco1}
\end{equation}
with $2 \lesssim \top \lesssim 6$.   \\

Now let us relate the mean density $\rho_{fco}$ and free-fall time $\tau_{ff,fco}$ of the failed cores to those of the host clump.  According to \citet{vzs05}, core-to-cloud mean density ratios of $\sim30\rm{-}100$ yield gravitationally unstable cores when these are modeled as Bonnor-Ebert spheres.  For our failed cores, we therefore adopt a weaker contrast $\mathcal{C} = 10$ between their density $\rho_{fco}$ and the host clump density $\rho_0$: 
\begin{equation}
\mathcal{C} = \frac{\rho_{fco}}{\rho_0}=10\;.
\label{eq:Cround}
\end{equation}

The \fft of failed cores then obeys $\tau_{ff,fco} = \tffi / \sqrt{\mathcal{C}}$, and we can modify Eq.~\ref{eq:tretfco1} as
\begin{equation}
t_{ret,fco} =  3.3 \top \frac{\tffi}{\sqrt{\mathcal{C}}}\;.
\label{eq:tretfco2}
\end{equation} 
With $\mathcal{C} \simeq 10$, the return time/re-expanding time of the failed cores becomes:
\begin{equation}
t_{ret,fco} \simeq  \top {\tffi} 
\label{eq:tretfco3}
\end{equation} 
with $2 \lesssim \top \lesssim 6$.   \\

\subsection{The ONC as a test-bed} \label{ssec:onc}

In Sec.~\ref{sec:gald}, we have made a preliminary estimate of the initial gas density and \fft of the ONC progenitor, based on the 1\,Myr \stf periodicity imposed by the observations.   We have found: $\rho_{clump}^{ONC} \simeq 10^3\,\Msppp$, hence $\tau_{ff,0}^{ONC} \simeq 0.26$\,Myr.  Does the corresponding $\top$-value make sense?  
Taking the 1-Myr time-span separating the successive stellar populations of the ONC as a proxy of the failed-core return time, we obtain $\top \simeq  t_{ret,fco} / \tffi \simeq 1\,{\rm Myr}/0.26\,{\rm Myr} \simeq 3.9$, which falls in the range of viable $\top$-values quoted above.  We conclude that the age-span between successive peaks of a stellar age distribution looks as a promising probe of the initial gas density.  A prerequisite for this is that the return time of failed cores in units of the clump free-fall time, namely, the $\top$ parameter (Eq.~\ref{eq:tretfco3}), is known.  This topic obviously deserves more scrutiny, observations as well as simulations. 
     
\begin{figure}
\begin{center}
\epsscale{1.10}  \plotone{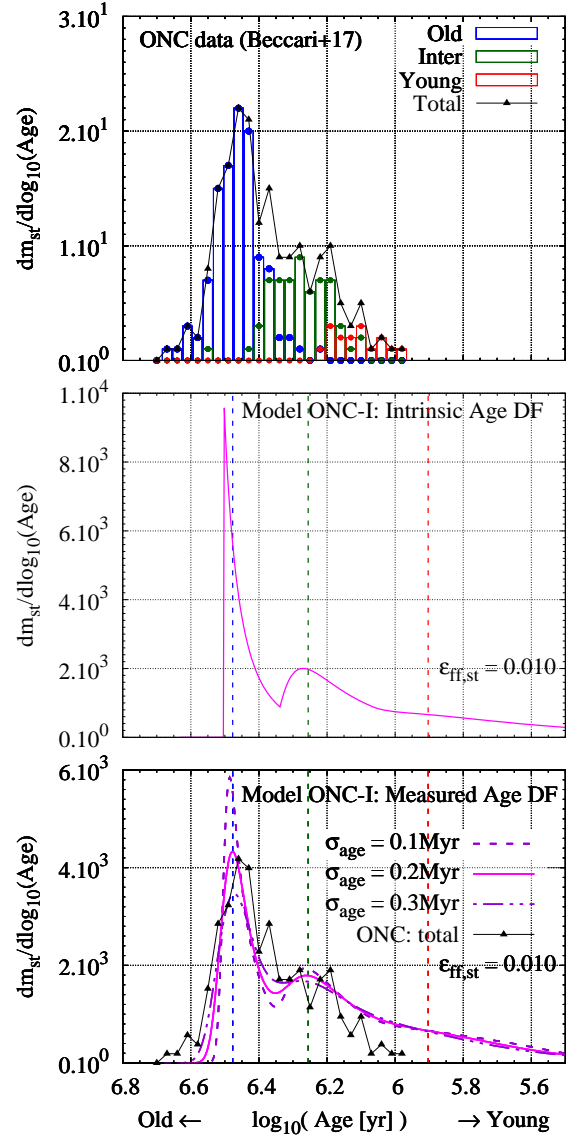}
\caption{Top panel: Logarithmic age distribution functions of the three populations of pre-main sequence stars identified by \citet{bec17} in the ONC (blue, green and red histograms; see key).  The sum of the three distributions is given by the black line with triangles.  Middle panel: Logarithmic stellar age distribution function for the ONC-I model (parameters given in Table \ref{tab:t3}).  The vertical blue, green and red dashed lines indicate the mean ages of the three populations as given in \citet{kro18}.  Bottom panel: Model age distribution function of the middle panel convolved by a Gaussian of standard deviation $\sigma_{age}$ accounting for stellar age uncertainties.  Results are presented for $\sigma_{age} =0.1, 0.2$ and $0.3$\,Myr (see key).  The solid black line with triangles is the ONC total age distribution from the top panel normalized so as to present the same old-peak amplitude as the model with $\sigma_{age} =0.2$\,Myr}
\label{fig:onc}
\end{center} 
\end{figure}   

Figure~\ref{fig:onc} compares the observed age distribution of ONC pre-main sequence stars with a model with initial gas density $\rho_0 = 10^3\,\Msppp$ and \cofe per \fft $\effco = 0.50$.
The top panel shows the age distributions of the three stellar populations identified by \citet{bec17}, using the same color-coding as in their Fig.~6.  Note that old ages are to the left and young ages (i.e. most recent star formation) are to the right.  The black line with triangles is the total age distribution.  Following the formation of the oldest population, the overall age distribution declines, highlighting a globally decreasing \sfr as found for other \sfing regions \citep[e.g. Chamaeleon~I, ][]{luh07,bel11b} and as predicted by \citet{par14} \citep[see also][]{prei12}.  We note that the three populations hardly show up when considering the total age distribution, an effect partly driven by stellar age uncertainties (see below).  

The middle panel presents the model age distribution for $\effco =0.5$, $\rho_0 = 10^3\,\Msppp$ ($m_{clump}=3.35\cdot10^4\,\Ms$ and $r_{clump}=2\,$pc), and an age of 3.3\,Myr for the onset of core formation.   Model parameters are quoted in Table \ref{tab:t3} under "ONC-I".  
The model age distribution presents two well-distinct peaks, prolonged by a tail of lower \stf activity.  The amplitude of the second peak is lower than that of the first/oldest one, as required to reproduce an overall decreasing \stf history.  The mean ages of the old, intermediate and young populations as given by \citet[][their Table 1]{bec17} are, respectively, 2.87, 1.88, and 1.24\,Myr, with the individual stellar ages taken from \citet{drio16}.  More recent estimates of the three population mean ages (based on a different set of pre-main-sequence isochrones) are mentioned in \citet[][their section 2.1]{kro18} as 3.0, 1.8 and 0.8\,Myr.  These three ages are highlighted as the blue, green, and red vertical dashed lines, respectively.  Note the shift of the youngest population towards even younger ages  (i.e. shift between the red histogram in the top panel and the dashed red line in the middle panel).  The most recent stellar population of the ONC would thus correspond to the trail of low-level \stf activity in the model of Fig.~\ref{fig:onc}.
 
 \begin{figure}
\begin{center}
\epsscale{1.10}  \plotone{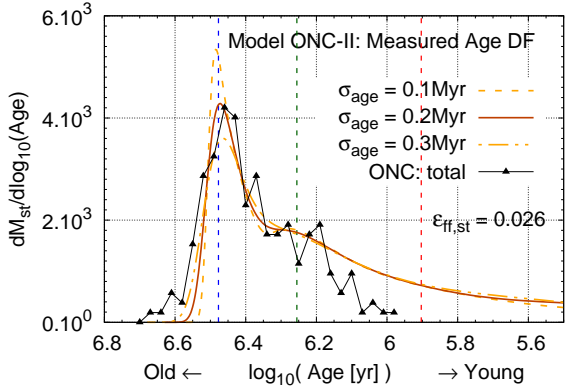}
\caption{Same as bottom panel of Fig.~\ref{fig:onc} but for the ONC-II model: $m_{clump}=10^4\,\Ms$, $r_{clump}=1.32$\,pc$, \effst=0.026$ and $\effco=0.30$ (see also Table \ref{tab:t3})}
\label{fig:onc_h}
\end{center} 
\end{figure}   

The bottom panel shows the impact of stellar age uncertainties.  For this, we convolve the model output with a Gaussian function of standard deviation $\sigma_{age}$, with $\sigma_{age}$ the uncertainty of individual stellar ages.  The mean  difference between the age estimates of the 3 populations as given by \citet{bec17} and \citet{kro18} amounts to 0.2\,Myr.  We thus adopt $\sigma_{age}=0.2$\,Myr.  At an age of 1\,Myr (3\,Myr), this equates with a logarithmic age uncertainty $\sigma_{log_{10}age}=0.09$ (0.03), in agreement with the uncertainties quoted in Table~5 of \citet{drio16}.    The convolved age distribution is given in the bottom panel of Fig.~\ref{fig:onc} as the solid magenta line.  We also show the result for a larger ($\sigma_{age}=0.3$\,Myr) or a smaller ($\sigma_{age}=0.1$\,Myr) uncertainty (dash-dotted and dashed purple lines).   At this stage, we refrain from a comprehensive browsing of the parameter space and from quantifying the fit quality with e.g. a $\chi^2$-test because of: (1)~the uncertainties of the observed stellar age distribution at young ages, and (2)~the absence of an initial gas density gradient in the current model.  How a density gradient will qualitatively affect our results is discussed later in this section and in Sec.~\ref{ssec:rhograd}.   

The relative amplitudes of the peaks of an {\it observed} star age distribution are of limited help to probe the history of a \sfing region.  This is because the peaks are unequally impacted by stellar age uncertainties.  For instance, in our model, the first/oldest peak is the sharpest, thus the one most curtailed by stellar age uncertainties.  A more robust indicator of the relative strengths of the successive \stf episodes is the stellar mass formed by each of them.  The three populations identified by \citet{bec17} contain, respectively,  111, 63 and 16 pre-main-sequence stars\footnote{\citet{bec17} mention a census of 24 stars for the ONC youngest population.  However, counting the number of objects in the red histogram of their Fig.~6 gives 16, and we use this latter number.}.  This is equivalent to 58\,\%, 33\,\% and 9\,\% of the ONC stellar mass if the mean stellar mass is the same for all three populations, and if the observed samples are representative of their respective populations.  
The model-predicted fractions are more even, however, with 40\,\%, 31.5\,\% et 28.5\,\% of the total stellar mass in the first, second and third million year, respectively.  Adding an initial density gradient to the model clump will contribute to alleviating this (see Sec.~\ref{ssec:rhograd}).   The total stellar mass produced by the model is $1400\,\Ms$, in rough agreement with that observed in the ONC \citep[$1800\,\Ms$, ][]{hh98}. \\

\begin{table}
\begin{center}
\caption{The ONC models of Sec.~\ref{ssec:onc}.  Common parameters are: $\rho_0=10^3\,\Msppp$, $t_{ret,fco}=1\,{\rm Myr}\equiv 3.9\,\tffi$, and an age of 3.3\,Myr for \cof onset.  Also given are the stellar mass and \stf efficiency at $t=3$\,Myr} 
\begin{tabular}{l c c c c c c c} \tableline 
Model            & ~$m_{clump}$~      & ~$\effco$~ &  $\effst$& ~$m_{st,3Myr}$~  & $SFE_{gl,3Myr}$ \\
 ID                 & ~[$10^4\Ms$]          &  ~~             &             & [$\Ms$]                  & \\ \tableline
 ONC-I  &    $3.3$   &             0.5          &    0.010   & $1400.$ & 0.04 \\
 ONC-II  &    $1.0$   &          0.3          &     0.026     &  $1500.$ & 0.15 \\ \tableline
\end{tabular}
\label{tab:t3}
\end{center}
\end{table}

The gas mass required by the ONC-I model may still seem prohibitive ($m_{clump}=3.35\cdot10^4\,\Ms$).  It is higher by a factor $\sim 3$ than any gas mass estimate proposed so far for the ONC progenitor \citep[e.g. $1.2\cdot10^4\,\Ms$; ][]{kro01}. 
One parameter which can still be adjusted to reduce the initial gas mass and whose variations have not been explored yet is the \sfe per free-fall time.  By increasing $\effst$, one can form a similar stellar mass out of a smaller amount of gas.  
\citet{pok21} infer, as an average for 12 nearby molecular clouds, $\effst = 0.026$.  A model with $\effst = 0.026$ and $\effco=0.30$ is shown in Fig.~\ref{fig:onc_h} (model ONC-II in Table~\ref{tab:t3}).  The clump mass and radius are, respectively, $m_{clump}=10^4\,\Ms$  and $r_{clump}=1.32$\,pc (the radius has been decreased so as to retain the same density $\rho_0$ as for the ONC-I model).  The contributions in stars, \sfing core envelopes and failed cores to the core mass budget are provided in Table~\ref{tab:t1}.  
The stellar mass formed after 3\,Myr is $m_{st}\simeq1500\,\Ms$, similar to that of the ONC-I model.  
The reason as to why a similar stellar mass forms out of a three-times smaller gas mass is two-fold here.  
Not only is the \sfe per free-fall time $\effst$ higher ($\effst=0.026$ vs. $\effst=0.01$), the \cofe per \fft $\effco$ is also  smaller ($\effst=0.30$ vs. $\effst=0.50$), which has been shown in Fig.~\ref{fig:2ts} to increase the clump \stf efficiency.  Note that this latter effect comes at the expense of reducing the trough between the old and intermediate-age populations (compare the magenta and brown solid lines in bottom panel of Fig.~\ref{fig:onc} and Fig.~\ref{fig:onc_h}, respectively).        

Having used the observed stellar age distribution and stellar mass of the ONC to constrain the mass, density and \cofe per \fft of its gaseous progenitor, it is tempting to compare our results with earlier studies.  \citet{kro01} and \citet{far20}  study the kinematics of the ONC stars, including those unbound by gas expulsion and stellar interactions.  The ONC model of \citet{kro01} builds on an initial gas mass of $1.1\cdot 10^4\,\Ms$ with a half-mass radius of $r_{0.5}=0.45$\,pc,  and a \sfe of 1/3.  It gives rise to a bound cluster which retains one third of its initial stellar content in the aftermath of gas expulsion and survives for about 150\,Myr.  Our ONC-II model has a similar initial gas mass, but is more extended, the half-mass radius of our uniform density clump being $1.32/2^{1/3} \simeq 1.05$\,pc.  A gas density gradient will make our model more compact, however.  For instance, the same gas mass ($m_{clump}=10^4\,\Ms$) distributed inside the same radius ($r_{clump}=1.32$\,pc) but according to a singular isothermal sphere yields an initial half-mass radius of $r_{clump}/2=0.66$\,pc for the gas.  The stars will be even more concentrated \citep{par13}.  Another difference between ONC-II and \citet{kro01} is the global \sfe $SFE_{gl}$.  \citet{kro01} assume $SFE_{gl}=0.33$ at gas expulsion, while our model predicts $SFE_{gl}=m_{st}/m_{clump}=0.15$ at $t=3$\,Myr.  Does our low \sfe imply that the ONC-II model cluster is unable to  survive gas expulsion?  Not once our model will include a gas density gradient.  Stars will then form more centrally-concentrated than the embedding gas \citep{par13}, and the locally higher \sfe achieved by its inner regions will help the cluster retain part of its stars despite a globally low \stf efficiency  \citep[see also][]{ada00}.  
Using $N$-body simulations building on the model of \citet{par13}, \citet{shu17} conclude that the \sfe threshold for cluster survival is $SFE_{gl}=0.15$, with instantaneous gas expulsion and an external tidal field (taken as that of the Solar neighborhood).  Not only will a gas density gradient help the cluster survive gas expulsion, it will also drive more efficient star formation \citep{par19}.  This will raise $SFE_{gl}$ above the threshold predicted by \citet{shu17}, and reduce the initial gas mass $m_{clump}$ accordingly (since the ONC stellar mass is imposed).      

\citet{far20} use their gradual cluster formation model \citep{far19} to model the velocity distribution function of the ONC stars, including run-aways and walk-aways.  They conclude that slow \stf ($\effst=0.01{\rm -}0.03$, rather than $\effst \geq 0.1$) and a progenitor clump surface density $\Sigma_{clump}\simeq 7000\,\Mspp$ (rather than $\Sigma_{clump}\simeq 700\,\Mspp$) provide the best match to the observations (their Fig.~10).   The surface density of our ONC-II clump $\Sigma_{clump}=10^4/(\pi 1.32^2) \simeq 1800\,\Mspp$ is 4 times lower than what they advocate.  Both models are not entirely comparable, however.  While our \stf histories are intrinsically time-varying, \citet{far19} assume a constant \stf rate fixed by the initial gas mass and \fft (their Eq.~7).  

Although, in principle, the comparison of different modeling approaches should help cast more light on the formation and evolution of star clusters, this is no straightforward task, and the variety of model assumptions calls for a careful comparison. 

\subsection{Do observed single-population clusters hide multiple star formation episodes ?}
\label{ssec:stburst}

\begin{figure}
\begin{center}
\epsscale{1.10}  \plotone{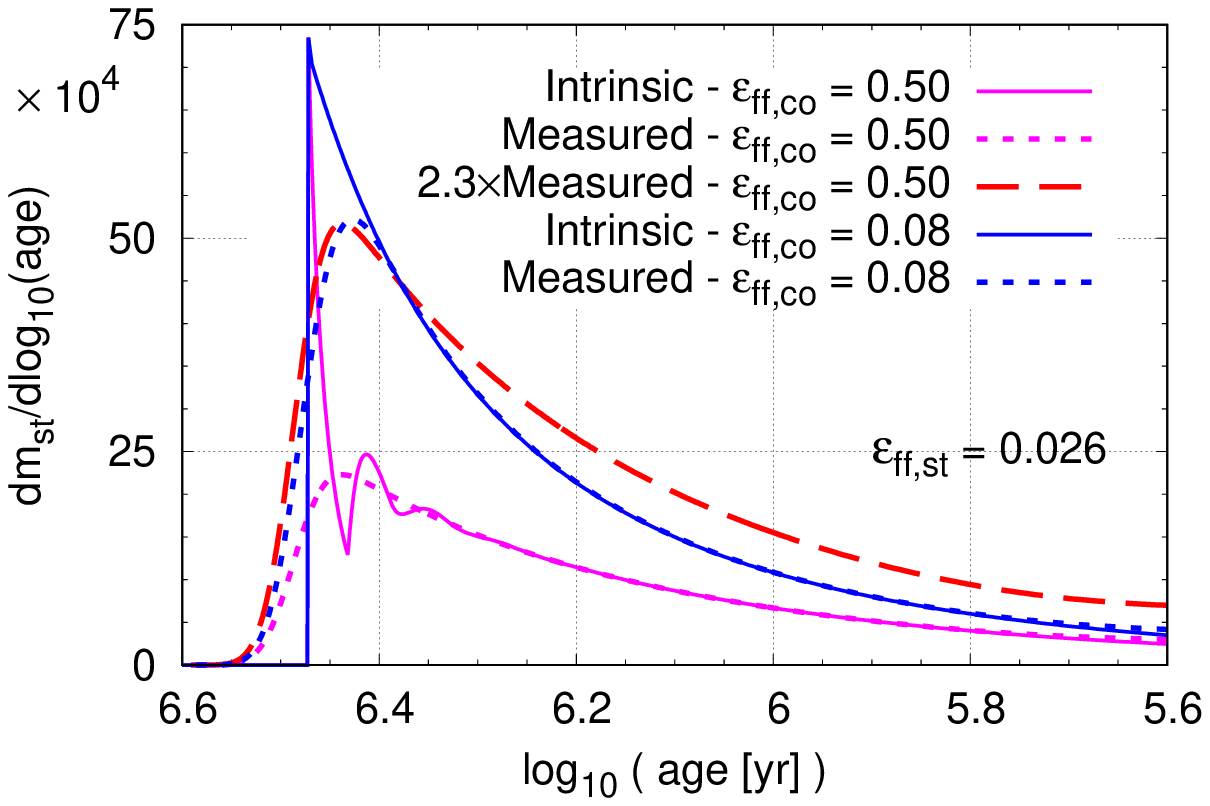}
\caption{Intrinsic (solid lines) and measured (dashed lines) stellar age distributions for a cluster-forming clump  with $\rho_0=10^4\,\Msppp$.  The blue and magenta lines correspond to $\effco=0.08$ and $\effco=0.5$, respectively (see key).  The red long-dashed line is the measured age distribution for $\effco=0.5$ raised to the same amplitude as for $\effco=0.08$: both measured age distributions are undistinguishable in shape}
\label{fig:burst}
\end{center} 
\end{figure}   

The frequency at which cluster-forming clumps pile up successive \stf episodes now depends on the clump density $\rho_0$ (Eqs \ref{eq:tretfco2} and \ref{eq:tretfco3}).  An interesting consequence is that the \stf histories of starburst clusters could have close-by peaks that, in the observed age distribution, get blended due to the short age-span separating them.  To illustrate this possibility, let us consider two clumps with the same mass $m_{clump}=3.3 \cdot 10^5\,\Ms$, radius $r_{clump}=2$\,pc (i.e. clump density $\rho_0 = 10^4\,\Msppp$) and \sfe per \fft $\effst = 0.026$, but distinct \cof efficiencies per free-fall time.  With $\effco=0.08$, the first clump form no failed cores ($F_{fco}=0.0$; see Table~\ref{tab:t1} for $\effst = 0.026$).  In contrast, with $\effco=0.5$, the second clump has a failed core mass fraction $F_{fco}=0.84$ (see Table~\ref{tab:t1}).  
Their respective intrinsic stellar age distributions (Fig.~\ref{fig:burst}) reflect this difference: a single spike when $\effco=0.08$, but 2-to-3 peaks for $\effco=0.50$.  Note also the lower amplitude of the stellar age distribution for $\effco=0.50$, which reflects the lower \sfe induced by higher $\effco$, as already noticed in Fig.~\ref{fig:2ts}.

The measured age distributions, that is, the intrinsic distributions convolved with a Gaussian error, tell a different story, however.  We adopt $\sigma_{age}=0.2$\,Myr, which is slightly smaller than what \citet{kud12} inferred for the Galactic starburst cluster Westerlund~1 ($\sigma_{age}=0.25$\,Myr).  Given the very short clump \fft $\tffi=0.08$\,Myr, the age span between two successive peaks is $\simeq 0.3$\,Myr ($\top =3.9$), and the multi-peak aspect shown by the $\effco=0.50$ model is erased once stellar age uncertainties are accounted for (see magenta short-dashed line in Fig.~\ref{fig:burst}).
Figure~\ref{fig:burst} also shows as the red long-dashed line the measured age distribution for $\effco=0.50$ normalized to the same maximum as for $\effco=0.08$.  The shapes of both measured age distributions are hardly  distinguishable, despite containing fundamentally different \stf histories.      

Starburst clusters in the Galactic disk - like Westerlund~1 - are observed to be single-burst clusters \citep{kud12}.  The exercise above shows that the actual underlying \stf history may be quite different, however.  We will revisit this issue in the future, once an initial gas density gradient will have been integrated into our model.  The reader may indeed ask as to why the full-width at half-maximum of our convolved age distributions in Fig.~\ref{fig:burst} is as large as $\simeq 1.5$\,Myr, that is, significantly larger than the observed full-width at half-maximum of $0.4$\,Myr inferred by \citet{kud12} for Westerlund~1.  This effect stems from the absence of an initial gas density gradient.  A density gradient $\rho_{gas} \propto r^{-2}$ will yield a full-width at half-maximum reduced by a factor 3-to-4 compared to a uniform density clump \citep[see Fig.~4 in][]{par19}, thereby bringing model and observed full-widths at half-maximum in agreement. 

\section{Discussion} \label{sec:disc} 
  
\subsection{Diversity in star-formation histories} \label{ssec:div}
Our model is able to generate very diverse \stf histories, from a \sfr steadily decreasing with time to 'jagged' \stf histories.  Additionally, the time-span between two successive peaks can vary from clump to clump, since it is a function of the initial gas density (Eq.~\ref{eq:tretfco2}).  

Figure \ref{fig:1myr} seems to suggest that more numerous \stf episodes arise in denser and/or more massive clumps (in each panel, compare red solid and magenta dashed lines).  This conclusion is erroneous, however.  In each panel of Fig.~\ref{fig:1myr}, the varying parameter is the clump mass, while the clump radius and the {\it physical} return time of the failed cores are kept constant.  Consequences are (1)~a modification of the clump density $\rho_0$, hence of the clump \fft $\tffi$ and, therefore, (2)~a modification of the return time of the failed cores in units of the clump free-fall time $t_{ret,fco}/\tffi$.  
The seemingly greater ability of more massive clumps to carve more peaks in their \stf history in Fig.~\ref{fig:1myr} stems from their failed core return time $t_{ret,fco}/\tffi$ becoming longer by virtue  of an increasingly shorter \fft $\tffi$ coupled to a {\it constant physical return time} $t_{ret,fco}$.  A longer return time of the failed cores $t_{ret,fco}/\tffi$ allows a greater decrease of the gas density by the time failed cores disperse, yielding thereby a greater relative increase of the gas density and of the \cofr at $t \gtrsim t_{ret,fco}$.  In other words, to give the first \cof episode more time to subside increases the likelihood of the second \cof episode to spike (see also Sec.~\ref{ssec:epi2} and Fig.~\ref{fig:S}).  Contrary to what a superficial glimpse at Fig.~\ref{fig:1myr} may suggest, the initial gas mass has no influence on the shape of the core and star formation histories.  

\begin{figure}
\begin{center}
\epsscale{1.10}  \plotone{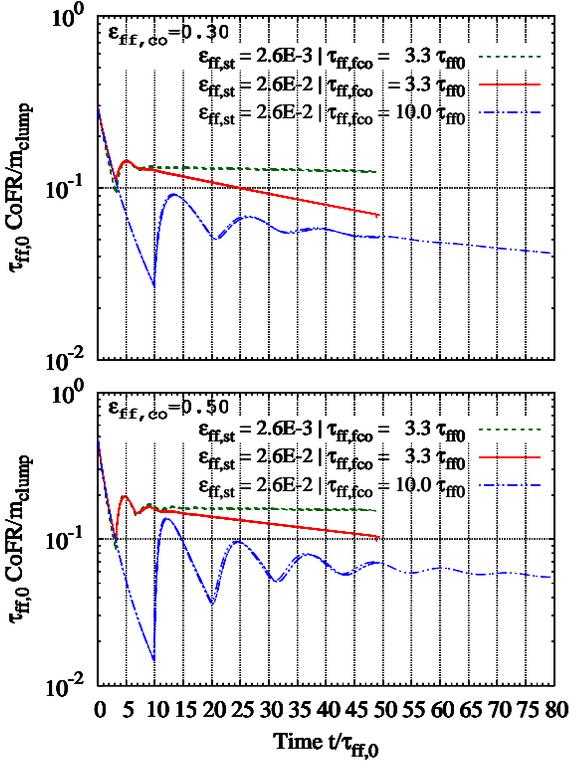}
\caption{Evolution of the \cofr normalized with the initial free-fall time $\tffi$ and initial mass $m_{clump}$ of the clump gas, as a function of time after core formation onset in units of the clump free-fall time.  Each line with given color and type corresponds to four model clumps with different masses $m_{clump}$ and densities $\rho_0$ (see text for details).} \label{fig:norm}
\end{center} 
\end{figure}   

To illustrate this even more convincingly, let us consider a clump of mass $m_{clump} = 3.3 \cdot 10^5\,\Ms$ and radius $r_{clump}=2$\,pc (density $\rho_0=10^4\,\Msppp$).  We also consider three clumps with a 10-times higher mass and either (i) the same radius ($m_{clump} = 3.3 \cdot 10^6\,\Ms$, $r_{clump}=2$\,pc, $\rho_0=10^5\,\Msppp$), or (ii) the same volume density ($m_{clump} = 3.3 \cdot 10^6\,\Ms$, $r_{clump}=4.3$\,pc, $\rho_0=10^4\,\Msppp$), or (iii) the same surface density ($m_{clump} = 3.3 \cdot 10^6\,\Ms$, $r_{clump}=6.3$\,pc, $\rho_0=0.3\cdot10^4\,\Msppp$).  

The mass $m_{clump}$ and the density $\rho_0$ define the {\it dimensional} parameters of the model.  For each pair ($m_{clump}$, $\rho_0$),  we vary the {\it dimensionless} model parameters, namely: the \cofe per \fft $\effco$, the return time of the failed cores in units of the clump \fft $t_{ret,fco}/\tffi$, and the \sfe per \fft $\effst$.  We test the following values: $\effco=(0.3, 0.5)$, $t_{ret,fco}/\tffi = (3.3, 10.0)$, and $\effst =(0.0026,0.026)$.  Twenty-four \cof histories are presented in Fig.~\ref{fig:norm} (top panel: $\effco=0.30$; bottom panel:  $\effco=0.50$).  Time on the $x$-axis is in units of the clump free-fall time, and the \cofr on the $y$-axis is normalized by the initial mass and \fft of the gas.  
Models with a given set of dimensionless parameters (i.e. lines with given color and dashed type in a given panel) display identical normalized \cof histories.  For instance, the 4 red solid lines, each depicting a ($m_{clump}$, $\rho_0$) combination, coincide with each other.  The normalized \cof history $\tffi CoFR/m_{clump}(t/\tffi)$ is therefore independent of the clump mass $m_{clump}$ and density $\rho_0$.  It is shaped by the parameter triple $(\effco, t_{ret,fco}/\tffi, \effst)$\footnote{For the sake of completeness, we should also mention the return time of the envelopes of \sfing cores $t_{ret,colco}/\tffi$.  However, given that \sfing cores represent a minor fraction of the total core mass in models generating successive \stf episodes, this fourth dimensionless parameter is here of minor importance.}
, while the clump mass and density (equivalently, the mass and free-fall time) act as scaling factors.  A larger clump mass raises the \cof history without modifying its overall shape.  A shorter \fft (i.e. denser clump) also provides a scaled-up version of the \cof history, while simultaneously increasing its "frequency".  

For the parameters tested, Fig.~\ref{fig:norm} shows that we can expect up to 5-6 \stf episodes (for $\effco=0.50$ and $t_{ret,fco}/\tffi=10$). 
The choice of $t_{ret,fco}/\tffi=10$ for the return time of the failed cores may seem extreme.  It requests $\top \simeq 10$, which goes beyond the range of values advocated in Sec.~\ref{ssec:tretfco} (see Eqs~\ref{eq:tretfco2}-\ref{eq:tretfco3}).  Yet, $t_{ret,fco}/\tffi=10$ may be relevant for environments where the external pressure is high enough to stabilize the cores  for a longer time-span than when it is negligible \citep{lad08}.   

Although the numbers of \stf episodes in Fig.~\ref{fig:norm} are comparable to the numbers of stellar populations identified in old \gcs \citep[often $\geq$3-4 subpopulations; e.g.][]{car15,bala18}, we note that the relation between the two is not that straightforward (see Sec.~\ref{ssec:gc})\footnote{An even greater number of stellar populations - 16 - have been extracted from the chromosome map of $\omega$~Cen, the most massive Galactic \gc \citep{mil17}.  $\omega$~Cen, however, is likely the nucleus of a now disrupted dwarf galaxy \citep[e.g.][]{hil20} and its early evolution cannot thus be explained with a cluster-based only scenario.  }.

The conclusion that the clump mass does not shape the \cof history seems at odd with what we know about old globular clusters, namely, that more massive globulars tend to present stronger multiple stellar population patterns.  For instance, the mass of Galactic \gcs correlates with their maximum helium abundance variation \citep{mil18}, with their maximum nitrogen abundance variation \citep{lag19}, and with the number fraction of stars enriched in hot hydrogen burning elements \citep{mil17}.  
Here, it is crucial to stress that the multiple stellar populations (generations if they fit a time sequence) of old \gcs are identified based on their chemical abundances.  In contrast, the successive \stf episodes of our models are identified based on the \stf history and the failed core return time $t_{ret,fco}$.  

The detailed \stf histories of old \gcs have remained unknown so far, and our model does not compound the \stf history of a cluster with its chemical evolution.
Yet, could our approach help shed light on as to why more massive \gcs present more prominent multiple-stellar-population patterns?

\subsection{Potential link with multiple stellar populations in old globular clusters} \label{ssec:gc}

Patterns of multiple stellar populations vary greatly among Galactic old globular clusters.  Specifically, the morphology of their chromosome map\footnote{Chromosome maps trace multiple stellar populations based on their star $UV$-photometry.  They build on the property that ultraviolet colors are very sensitive to star-to-star helium and light-element abundances \citep{mil15}}
is highly diverse - with no one cluster being identical to another.  For instance, the branch of the so-called second generation is sometimes more clumpy (e.g. NGC2808), sometimes more continuous \citep[e.g. NGC6652; see Fig.~4 in][]{mil17}.  These maps also reveal varying numbers of globular cluster subpopulations.   Similar conclusions can be  drawn from the sodium-oxygen anticorrelation \citep[see e.g. Fig.~1 in][]{cha16}.  
It is tempting to connect this aspect to our model \stf histories, whose aspect is sometimes 'wavy', with well-defined peaks and troughs (when $\effco$ and $t_{retfco}/\tffi$ are large), sometimes smoothly decreasing with time.  But first, we have to cope with the apparent contradiction noted in Sec.~\ref{ssec:div}, that high-mass globular clusters present the most striking patterns of multiple stellar populations and of enrichment in \hhb elements, while the shape of our model \stf histories are independent of the progenitor clump mass (Fig.~\ref{fig:norm}).  Can we get out of this predicament?  

Yes, if we assume that the stellar mass built by the clump must reach a given threshold $m_{th}$ to trigger the formation of the polluters providing the hot hydrogen burning elements.  This could be the case if the polluters are supermassive stars, namely, stars with masses  a few $10^3$ - a few $10^4\,\Ms$ \citep{den14}.  Their predicted yields are in excellent agreement with the light-element abundance patterns observed in \gcs (e.g. the sodium-oxygen anti-correlation: compare Fig.~1 in \citet{cha16} with the top-right panel of Fig.~1 in \citet{den14}).  In addition, \smss could pollute their environment during a few million years \citep{den14}, similar to the time-span for cluster development.  Their formation mechanism would be via stellar collisions, which requires either a high cluster density \citep[e.g.][]{spz04}, or a high number of stars \cite[at least a few $10^6$ stars;][]{gie18}.  Here, we build on the scenario that their formation requires a minimum number of stars, hence a minimum stellar mass $m_{th}$.

Let us consider three gas clumps (1,2,3) of mass $m_1<m_2<m_3$.  The least massive clump is also less massive than the threshold mass $m_{th}$.  With $m_1<m_{th}$, clump~1 is unable to form a supermassive star, and the corresponding cluster does not display the light-element abundance variations typical of globular clusters.  Yet, if the \cofe $\effco$ and return time $t_{ff,fco}/\tffi$ of clump~1 are large enough, its \stf history does display successive \stf episodes.  
This may explain why open clusters do not present light-element abundance variations \citep{mcl15}, despite the fact that some (future) open clusters present multiple \stf episodes \citep[e.g. the ONC, ][]{bec17}.  

We now consider clump~2 and clump~3 as massive enough to form a supermassive star, i.e. $m_{th}<<m_2<m_3$.  We also assume that their mass is the only parameter differentiating them (i.e. their free-fall times are equal, as well as their dimensionless parameters).  With $SFE_{2,th}$ and $SFE_{3,th}$ the \stf efficiencies of clump~2 and clump~3 at the time they reach the stellar mass threshold $m_{th}$, we have $SFE_{3,th}<SFE_{2,th}$ since $m_{th} = SFE_{3,th} m_3 = SFE_{2,th} m_2$.  Because both clumps have identical free-fall times and dimensionless parameters, the growths with time of their respective \stf efficiencies are alike, $SFE_2(t)=SFE_3(t)$.  Therefore, the inequality $SFE_{3,th}<SFE_{2,th}$ implies that clump~3 reaches the threshold $m_{th}$ {\it at an earlier time} than clump~2.  This shift in time brings forward the formation of a supermassive star in clump~3, thereby providing a longer pollution time-span.  As a result, the more massive clump should form a larger fraction of polluted stars, and achieve more extreme light-element abundance variations, as observed \citep{mil17,mil18,lag19}.

As long as the supermassive star does not "operate", newly-formed stars constitute the - chemically-defined - first stellar generation.   
We emphasize that the end of this generation coincides with the onset of the supermassive star pollution, not with the end of the first \stf episode of our model.  If the \icm pollution starts during the third \stf episode, the first, second and beginning of third \stf episodes define the first stellar generation.  
Conversely, one could consider the case of a protoglobular clump with $\effco < 0.1$, but massive enough to form a supermassive star. 
Its \stf history would be peakless (small \cofe per free-fall time, hence small return fraction $f_{ret,fco}$).  Yet, the chromosome map would present two groups of stars corresponding to the stars born before and after the onset of the supermassive star pollution.  In other words, the shape of the \stf history and the chemical evolution are decoupled from each other.  While the shape of the \stf history depends on the model {\it dimensionless} parameters, the chemical evolution depends on a {\it dimensional} parameter (in the present case, the threshold stellar mass $m_{th}$).  This is the very reason as to why, all through this paper, we have refrained from using the term "stellar generations" to describe the successive \stf episodes of our model, this term being often used in studies of \gc multiple stellar populations.  A consistent comparison between the number of \stf episodes predicted by our model and the number of multiple stellar generations in old \gcs calls for a comparison between the return time of the failed cores with the  chemical enrichment time-scale, which is beyond the scope of the present study. \\

Finally, we stress that the early evolution of \gcs may see successive pollution episodes, driven by different polluters operating on different time-scales \citep[e.g.][]{dant16}.  Despite its low-mass \citep[$2.2\cdot10^4\,\Ms$, ][]{bau19}, the \gc NGC6535 hosts two chemically-defined stellar populations \citep{mil17}.  How could such a low-mass \gc contain stars bearing the imprint of \hhb ?  Either the pollution of protoglobular clumps in \hhb products is not caused by \smss only.  Or the current mass of NGC6535 is a tiny residual of a much larger initial mass.  In either case, it is interesting to note that NGC6535 hosts two chemically-defined populations, while the much less massive ONC presents three \stf episodes.  One more evidence that peaks in \stf histories are not direct matches to \gc multiple stellar generations.  

\subsection{Caveats and Limitations}
\label{ssec:modlim}

Our idealized model does obviously not capture all aspects of the \stf process.  In this section we assess its limitations and how more realistic conditions could affect its outcomes.  We start by discussing our assumption of an isolated clump in near-equilibrium (i.e. constant clump mass and radius).
  
The constancy of the clump radius constitutes a model central hypothesis since the gas density and \cofr depend sensitively on it.  Under what conditions can a cluster-forming clump remain in near-equilibrium once \stf has started?  A key ingredient here is the feedback from forming stars, which stir their surrounding gas.  Protostellar winds given off by low-mass stars were already suggested by \citet{nor80} to be prime drivers of turbulence within molecular clouds.  On the pc-scale of cluster-forming clumps, \citet{nak07} reach the same conclusion by means of hydrodynamical simulations.  They show that even in case of fast turbulence decay, turbulence gets replenished from inside the clump due to the gas motions generated by protostellar outflows.  This happens because protostellar outflows are collimated \citep[i.e bipolar outflows; see][for a review]{bal16}, allowing them to transport energy and momentum to greater distances from their sources than if they were spherically-symmetric \citep{nak07}.  This "protostellar outflow-driven turbulence"  allows the model clump of \citet{nak07} to remain close to dynamical equilibrium provided that the \stf efficiency per \fft amounts to a few per cent.  There is in fact an interplay between the rates of turbulence decay and star formation.  \citet{mat01} shows that a near-equilibrium configuration arises when turbulence decays slowly (over 10 free-fall times).  When turbulence decays fast (over one free-fall time), however, continuous \stf is required for protostellar outflows  to offset turbulence decay.  If \stf happens discretely (i.e. one star at a time due to a low \stf rate), turbulence decay cannot be offset and the system collapses\footnote{An example of a cluster-forming clump whose shrinking is unimpeded by protostellar outflows is provided in \citet{huf06}.}.  The \sfe per \fft of our models ($\effst = 1$-$2.6$\,\%) is close to the limit for turbulence replenishment inferred by \citet{nak07} and, therefore, our assumption of a constant clump radius appears reasonable.  Additionally, we note that even if our model had to include the contraction of a clump, this contraction is more likely to happen when the \sfr reaches a low point, that is, one return time after a peak in the \stf history.  Such a clump contraction would be timely to stimulate even more the rise of the next \stf episode by increasing the clump gas density. 
 
One could now ask what the fate of these protostellar outflows is.  Do they leave the cluster-forming clump \citep[e.g.][]{far19}, thereby decreasing the gas mass available to star formation, or do they remain trapped in the clump (our model).  That is, how valid is the assumption of an isolated clump?
This is a complex issue since it involves (1) the outflow speed and the clump escape velocity (hence the clump mass, radius, density profile, and the location of the source star inside the clump), and (2) the outflow morphology \citep[which often consists of an inner high-velocity jet, superimposed to a slower wider-angle component,][]{bal16}.  The escaping outflows can also sweep part of the clump gas on their way out, thereby increasing the gas mass fraction lost from the clump.  \citet{fed14} find that more than a quarter of the clump mass can be ejected this way.  On the one hand, the \sfr would subside more quickly than predicted by our model, as a result of the corresponding gas mass decrease.   On the other hand, additional star formation may be triggered since jets and outflows also drive local turbulent gas compressions \citep{fed14}.  Should the formation of failed cores be enhanced too, the multiple stellar population patterns predicted by our model will be strengthened.
We also note that, when the leftover \sfing gas is completely expelled out of protoclusters due to massive star feedback,   \stf may be triggered in the ensuing shell of expanding gas.  This will add to the phenomenon of star cluster multiple stellar populations if the stellar component of the shell collapses back \citep{par04,rah18}.  \\

Another way of modifying the mass enclosed within the adopted clump radius is by gas inflows.  \citet{mye09} find that all nine young stellar groups within 300\,pc of the Sun radiate multiple filaments of parsec length, designing thereby a hub-filament structure.  Gas flowing inwards along these filaments can fuel the cluster \stf process.  This may be the case for the Serpens South embedded cluster \citep{kir13} and for the ONC \citep{hac17a}.  In the later case, \citet{hac17a} estimate that the integral-shaped filament currently supplies the ONC with gas at a rate of $\dot{M}_{acc} \simeq 385\,\MsMy$.  This accretion rate cannot impact significantly the clump gas reservoir of our ONC-II model whose initial gas mass is $10^4\,\Ms$.  This conclusion, however, would be different if the gas accretion rate was higher in the past \citep{kro18}.  To assess the importance of filamentary accretion onto forming clusters cannot be performed without knowledge of the gas accretion rate history.  In the case of our model, the effect would be to fill in the troughs in the time evolution of the gas mass and, therefore, to reduce the amplitude of the oscillatory patterns seen in some of our simulations, which we now turn to.    \\

Our model \stf histories can present clear-cut oscillations, which happens when failed cores dominate the core mass budget.  These oscillations stem from the adoption, for each model clump, of a unique return time for all failed cores, i.e. $t_{ret,fco} =  3.3 \top \tffi / \sqrt{\mathcal{C}}$ (with $3.3\top$ the return time of the failed cores in unit of their free-fall time, and $\mathcal{C}$ the core-to-clump mean density ratio; see Eqs~\ref{eq:tretfco2}, \ref{eq:Cround}, \ref{eq:tretfco1}).  The amplitude of the oscillations as predicted by our model therefore constitutes an upper limit: in case of spatial variations in the core return time, the gas replenishment from simultaneously-formed cores is spread more evenly in time, smoothing thereby model oscillations.  The effect would be especially significant for cores formed at a peak of core formation activity, since these cores bear  the subsequent  \cof history "upheaval".  Although we have assumed that the clump gas remains uniform in density (albeit declining with time) through the clump evolution, it is not the case in practice.  Hydrodynamical simulations of turbulent self-gravitating clouds produce filaments \citep[e.g.][]{fed16,smi16}, implying that the immediate environment of cores varies from one core to another, an effect also reinforced by spatial variations of the ram pressure arising from protostellar outflows \citep{nor80}.  With re-expanding cores confined by different external pressures, the $\top$ parameter and, therefore, the return time $t_{ret,fco}$ must vary from core to core.  The mean density contrast $\mathcal{C}$ of failed cores can also vary (although remaining markedly smaller than that of \sfing ones).  More realistic simulations may thus show that the formation of clearly distinct \stf episodes is the exception rather than the rule, even when failed cores dominate the core mass budget by far.  Interestingly, inspection of figures 3-7 in \citet{mil17} suggests that \gcs whose enriched population draws a smooth distribution of points in the chromosome map outnumber those with clearly clumpy distributions  (although photometric uncertainties contribute to this as well).  \\

Another aspect that our present model does not account for is that of core coalescence \citep[e.g.][]{nor80,dib10}.  The corresponding mass increase may allow some of them to reach their Jeans mass, thereby turning failed cores into \sfing ones.  This too would reduce the amplitude of the \stf history oscillations. 

Coalescence and return-time variations of failed cores contribute to explaining why the strong oscillatory patterns observed in some of our models do not show up in hydrodynamical simulations\footnote{At least we are not aware of such simulations.  Unfortunately, such publications do not always present their star or core formation history.}: these are characterized by many more degrees of freedom than our model.  But we also stress that such simulations need to cover at least 4-6 free-fall times of the modeled \sfing region for two subsequent oscillations to be uncovered.

Finally, a note about magnetic fields.  The higher their initial strength in the cluster-forming clump, the lower the \cofe per free-fall time \citep{dib10}.  Magnetic field differences among cluster-forming clumps may thus contribute to whether or not subsequent \stf episodes emerge in  clusters.

\subsection{Impact of an initial gas density gradient}
\label{ssec:rhograd}
In this contribution, we have restricted our attention to uniform density clumps.  Yet, observed cluster-forming clumps are centrally-concentrated \citep[e.g.][]{mue02}.  Compared to a top-hat profile, a gas density gradient strengthens the \stf activity of a clump by locating a greater fraction of its gas mass in its central regions, where the gas density is higher and the \fft shorter.   To quantify this effect, \citet{par19} introduces the magnification factor $\zeta$, defined as the ratio between the \sfr of a centrally-concentrated clump and that of its top-hat equivalent\footnote{The top-hat equivalent is  a clump of uniform gas density containing the same gas mass as the centrally-concentrated clump, enclosed within the same radius.} \citep[her equations 8-9; see also][]{tan06}. 
For instance, a gas density profile with a logarithmic slope of $-2$ \citep{mue02} and a central plateau of extent $10^{-2}$ the clump radius yields $\zeta \simeq 2.5$ \citep[see top panel in Fig.~1 of][]{pp20}.  
The higher the gas density contrast between the clump center and the clump edge, the higher the magnification factor is, and the more invigorated \stf activity is. 

In the framework of the present model, how would a density gradient affect our results?
The \sfr of a centrally-concentrated clump is given by Eq.~8 in \citet{par19}, where $\effint$ is the intrinsic \sfe per free-fall time, namely, the efficiency of the clump top-hat equivalent.  The same equation can be rewritten as a \cof rate, by substituting the \cofe per free-fall time, $\effco$, for the \sfe per \fft $\effint$:
\begin{equation}
CoFR_{clump} = \zeta \effco \frac{m_{gas}}{\tff}\;.
\end{equation}   
In this equation, the magnification factor $\zeta$ can be associated either to the intrinsic \cofe per \fft $\effco$ \begin{equation}
CoFR_{clump} = (\zeta \effco) \frac{m_{gas}}{\tff}\,,
\end{equation}   
or to the clump gas mass
\begin{equation}
CoFR_{clump} \propto \effco \left(\frac{\zeta^{2/3} m_{gas}}{r_{clump}}\right)^{3/2}\;.
\end{equation}   

In the first case, the magnification factor raises the efficiency per free-fall time, implying that an initial gas density gradient strengthens the emergence of successive \stf episodes.  In the second case, a given \cof rate can be achieved with a gas mass reduced by a factor $\zeta^{2/3}$.  Equivalently, the clump \sfe $SFE_{gl}$ increases by a factor $\zeta^{2/3}$.  For instance, the \sfe of the ONC-II model ($SFE_{gl}=0.15$, Sec.~\ref{ssec:onc}, Table~\ref{tab:t3}) will be increased by a factor  $1.6{\rm -}2.1$ if $\zeta = 2{\rm -}3$. 

We also expect a gas density gradient to modify the distribution of the stellar mass among the successive \stf episodes.  The "weight" of the first/oldest \stf episode will increase at the expense of the subsequent ones, since the gas density contrast from clump center to clump edge plummets with time, causing in turn  a decrease of the magnification factor \citep[see Figs~2, 5 and 6 in][]{par19}.  

We will investigate these aspects in future papers.

\section{Conclusions}\label{sec:conc}

We have proposed a novel approach to understand the multiple stellar populations of star clusters.  Our model hinges on the presence of failed cores in \sfing regions.  Failed cores are density peaks of the interstellar medium which do not end collapsing and forming stars.    Instead, they re-expand and disperse back in the inter-core gas.  Although their existence has been known for more than two decades, both in observations \citep{tay96,lad08,bel11a,bel11b} and simulations \citep{cla05,vzs05,gom07}, they have not received the same scrutiny as their star-forming counterparts.  
The formation of failed cores necessarily depletes the gas reservoir of a \sfing region.  We therefore propose that their formation prompt the gas density, core and \stf rates to dwindle.  Their subsequent dispersion, however, allows the density of the surrounding gas to re-increase, thereby rejuvenating the core and star formation rates.  When failed cores dominate the total core mass, this controlling mechanism of the available gas mass and density yields peaks and troughs in the time evolution of the gas density, hence of the core and \stf histories.  The time-span on which failed cores disperse - which we coin their return time - determines the age-span separating the  successive peaks of a \stf history.  

Multiple peaks in the stellar age distribution of cluster-forming clumps \citep[e.g. the ONC;][]{bec17} imply that a \stf episode must at some point recede, to allow the next one to rise.  To implement the scenario described above, 
we therefore build on the volume-density driven cluster-formation model of \citet{par13}, which predicts a declining \stf activity in cluster-forming clumps under the assumptions of constant clump mass and size. 
 
We start in Sec.~\ref{sec:modI} with a preliminary model in which \sfing and failed cores release their gas on the same time-scale.  We find that three dimensionless parameters determine whether a \cof history unfolds at least two peaks: the \cofe per free-fall time $\effco$, the fraction $\fret$ of the mass in cores that returns to the clump gas, and the associated time-span $\tret/\tffi$, this one being expressed in units of the clump free-fall time $\tffi$.  High \cof efficiencies per \fft $\effco$, high return fractions $\fret$, and  long return times $\tret/\tffi$ favor the rise of a second \cof episode (see red-to-yellow areas in Fig.~\ref{fig:S}).    

Following a review of the evidence for failed cores in observed and simulated \sfing regions in Sec.~\ref{sec:fco}, we start to augment our model in Sec.~\ref{sec:modII}.  First, we introduce the \sfe per \fft $\effst$, whose combination with the \cofe per free-fall time $\effco$ defines  the respective fractions of the core mass turned into stars ($F_*$), given off to the clump gas as \sfing core envelopes ($F_{env}$), and as failed/rebounding cores ($F_{fco}$) (see Eqs~\ref{eq:mf1}-\ref{eq:mf3}, and Table~\ref{tab:t1}).  Secondly, star-forming cores lose their envelope on a shorter time-span than failed cores disperse.  We thus lift our assumption of identical return times for the failed cores and the \sfing core envelopes, and we assign them different time-scales (Eqs~\ref{eq:tretcolco}-\ref{eq:tretf}).  We assume that failed cores disperse on a time-span $t_{ret,fco}=1$\,Myr, which is the estimated life-expectancy of failed cores in the Pipe molecular cloud \citep{lad08}, and the core re-expansion time in the magnetohydrodynamical simulations of \citet{vzs05}.  This is also the age-span separating the successive populations of pre-main sequence stars in the ONC \citep{bec17},  

Next, in Sec.~\ref{sec:gald}, we generate stellar age distributions for various clump masses and  \cof efficiencies per \fft  (Fig.~\ref{fig:1myr}).  All model clumps are given the same \sfe per \fft $\effst$, radius $r_{clump}$, and physical return time  $t_{ret,fco}=1$\,Myr.  Based on the number of peaks being carved in the \stf history with the imposed 1-Myr-periodicity and the corresponding stellar mass, we estimate that the ONC gaseous progenitor had a mean gas density $\rho_0=10^3\,\Msppp$ and a \cofe per \fft $\effco \geq 0.3$.  
It bears repeating that this is the long return time of the failed cores $t_{ret,fco}/\tffi$ which drives the 'jagged' structure of the \stf history of the most massive clumps, {\it not} the clump mass itself.    

In Sec.~\ref{sec:ipts}, we lift the assumption of a constant physical return time $t_{ret,fco}$ for the failed cores.  
Using the analytical estimate of \citet[][see left panel of their Fig.~1]{vzs05}, we infer that the re-expanding time of failed cores amounts to a few free-fall times of their host clump (Eq.~\ref{eq:tretfco3}, with the \fft defined as in Eq.~\ref{eq:tff}).  We also propose two models for the \stf history of the ONC (Figs~\ref{fig:onc}-\ref{fig:onc_h} and Table~\ref{tab:t3}),  and emphasize that the \stf history of starburst clusters may contain several \stf episodes, these remaining concealed (i.e. blended) due to their inter-peak age-span being comparable to stellar age uncertainties.  

In Sec.~\ref{sec:disc}, we discuss various aspects of our model.  First, we emphasize the great diversity in the core and \stf histories we have generated.  The {\it shapes} of the core and \stf histories depend on the dimensionless model parameters ($t_{ret,fco}/\tffi$, $\effco$, $\effst$).    
That is, the evolution of the normalized \cofr $\tffi CoFR / m_{clump}$ as a function of time $t/\tffi$ depends exclusively on ($t_{ret,fco}/\tffi$, $\effco$, $\effst$) (Fig.~\ref{fig:norm}).  The {\it dimensional} parameters ($m_{clump}$, $\rho_0$) (i.e. clump mass and density) act as normalizing factors.  Second, we discuss the potential relation between our model and the multiple stellar populations/generations ubiquitously observed in Galactic old globular clusters.  That more massive \gcs present stronger patterns of multiple stellar populations seems, at first sight, to contradict our finding of mass-independent \stf histories.  Our \stf episodes, however, are defined based on the failed core return time, while the multiple stellar generations of old \gcs are mapped based on their light-element abundances (e.g. chromosome map, \citet{mil15}, and sodium-oxygen anti-correlation, \citet{car10}).  We qualitatively show that our model can produce massive clusters with a greater fraction of enriched stars if the formation of the stars polluting the cluster with \hhb elements requires a threshold stellar mass.  Third, we discuss the caveats and limitations of our model, emphasizing that more realistic conditions will smooth the strong oscillatory patterns carved in some of the predicted \stf histories.  Finally, we discuss how the inclusion of a density gradient in our model clumps will impact our present results.  We will chart in greater depth these additional aspects, along with that of multiple stellar populations, in future papers.  

\acknowledgments

GP acknowledges funding by the Deutsche Forschungsgemeinschaft (DFG, German Research Foundation -- Project-ID 138713538 -- SFB 881 ("The Milky Way System", subproject B05).  This research has made use of NASA 's Astrophysics Data System.  We thank the anonymous referee for a constructive report.

\end{document}